\documentclass[prd,twocolumn,superscriptaddress,floatfix,amsmath,amssymb,amsfonts,nofootinbib,longbibliography]{revtex4-2}

\usepackage{float} 
\usepackage{scalerel}
\usepackage[normalem]{ulem}
\usepackage[english]{babel}
\usepackage{graphicx}
\usepackage{dcolumn}
\usepackage{bm}
\usepackage{blindtext}
\usepackage{verbatim}
\usepackage{relsize}
\usepackage{mathrsfs}
\usepackage{musicography}
\usepackage{amsmath}
\usepackage{blindtext}
\usepackage{cancel}
\usepackage{physics}
\usepackage{epstopdf}
\usepackage{mathtools}
\usepackage{blindtext}
\usepackage{tensor}
\usepackage{color}
\usepackage[usenames,dvipsnames]{pstricks}
\usepackage{epsfig}
\usepackage{pst-grad} 
\usepackage{pst-plot} 
\usepackage{hyperref}
\usepackage{verbatim}
\usepackage{slashed}
\usepackage{dsfont}
\usepackage[english]{babel}
\usepackage{amsmath,amssymb,amsthm}  
\usepackage{tikz}
\usepackage{subcaption}

\newcommand{\mf}{\mathsf}

\newcommand{\ii}{\mathrm{i}}

\newcommand{\tbb}[1]{#1}  

\allowdisplaybreaks[1] 

\begin{document}

\title{The Einstein Tensor from UV-regulated Field Correlations}

\author{Matheus H. Zambianco}
\email{mhzambia@uwaterloo.ca}

\affiliation{Department of Applied Mathematics, University of Waterloo, Waterloo, Ontario, N2L 3G1, Canada}
\affiliation{Institute for Quantum Computing, University of Waterloo, Waterloo, Ontario, N2L 3G1, Canada}
\affiliation{Perimeter Institute for Theoretical Physics, Waterloo, Ontario, N2L 2Y5, Canada}

\author{Achim Kempf}
\email{akempf@uwaterloo.ca}

\affiliation{Department of Applied Mathematics, University of Waterloo, Waterloo, Ontario, N2L 3G1, Canada}
\affiliation{Institute for Quantum Computing, University of Waterloo, Waterloo, Ontario, N2L 3G1, Canada}
\affiliation{Perimeter Institute for Theoretical Physics, Waterloo, Ontario, N2L 2Y5, Canada}

\begin{abstract}

It is known that the spacetime metric can be recovered from quantum field correlators. We extend this programme by showing that the Einstein tensor can be directly written in terms of UV-regulated field correlations. The same correlation structure also directly determines the Ricci tensor, while higher coincidence derivatives recover the connection and Riemann tensor. We further show that the renormalized stress-energy tensor can be formulated within this framework, up to the standard finite renormalization freedom. These results provide a correlation-based organization of the semiclassical Einstein equations and advance the programme of formulating spacetime geometry in terms of quantum field correlations.

\end{abstract}

\maketitle

\section{Introduction}
\label{sec:introduction}

Quantum field theory (QFT) in curved spacetimes provides the standard
framework for describing quantum fields in the presence of a classical
gravitational background and has led to many advances in our understanding
of regimes in which quantum and gravitational effects are simultaneously
relevant~\cite{BirrellDavies,WaldQFTCS,ParkerToms}. Among its best-known
predictions are particle creation in expanding cosmological
spacetimes~\cite{ParkerOG,Parker1968}, the thermal properties of black
holes~\cite{Bekenstein1973,BardeenCarterHawking1973,Hawking1975,HawkingRadiation},
the observer-dependent particle content revealed by the Unruh
effect~\cite{Fulling1973,Unruh1976,matsasUnruh}, and the thermodynamics of
cosmological horizons~\cite{GibbonsHawking_1977}. When the backreaction of
quantum matter on the geometry is taken into account, the corresponding
effective description is provided by semiclassical gravity, in which the
classical Einstein tensor is sourced by the renormalized expectation value
of the quantum stress-energy tensor~\cite{Wald1977BackReaction,Wald1978TraceAnomaly,bibleHadamard}.

In these frameworks, spacetime geometry is introduced as an
independent structure on which quantum fields propagate. However, the
correlation functions of quantum fields themselves contain detailed
information about the underlying geometry. More concretely, for a class of
physically admissible states, the ultraviolet (UV) structure of the
two-point function of a scalar quantum field is constrained by the Hadamard
condition~\cite{fullingHadamard,fullingHadamard2,bibleHadamard,fewsterNecessityHadamard,Radzikowski_1996}.
\tbb{This condition implements the expectation that, at
sufficiently short distances, the leading singularity has the local
Minkowski vacuum form. It also allows the definition of renormalized
local observables, including the stress-energy tensor, by subtracting
a universal singular part determined by spacetime geometry and field
equation. These properties motivate its use as a criterion of physical
admissibility~\cite{bibleHadamard,fewsterNecessityHadamard}.}
In particular, the geometric singular part is governed by Synge's world
function, which is one half of the squared geodesic distance between
sufficiently nearby spacetime points~\cite{poissonbible}.

Consequently, the short-distance structure of correlations between sufficiently nearby spacetime points carries information about their geodesic separation. Building on this observation, the pioneering work of Ref.~\cite{Saravani_2016} showed that the spacetime metric can be reconstructed from quantum field correlators. This construction was subsequently extended to the Wightman function, and an operational interpretation in terms of localized measurements of field correlations was developed in Ref.~\cite{Perche_Edu_geo_2022}. Moreover, curvature can be inferred from the excitation probability of a particle detector ultra-rapidly coupled to a quantum field~\cite{Perche_Ahmed_2022}, and it also leaves observable imprints on localized field observables~\cite{AhmedMatheusTales2025}. Together, these results support the broader viewpoint that the local notion of spacetime distance need not be regarded as primitive but can, in principle, be inferred from the short-distance structure of quantum field correlations~\cite{Kempf_2021}.

This observation raises a natural question. If the metric is encoded in a
two-point function, to what extent can Einstein's equations themselves be
written in terms of field correlations? At the purely geometric level, the
existing metric-recovery formulas~\cite{Saravani_2016, Perche_Edu_geo_2022} already contain enough information to
answer this question indirectly. Once the metric has been reconstructed on
an open neighbourhood, one may calculate its derivatives, construct the
connection and curvature tensors, and then obtain the Einstein
tensor. 

\tbb{
This indirect procedure uses the standard
relation between the metric and the Einstein tensor.
We instead ask whether the tensor entering the geometric
side of Einstein's equations can be identified directly
within the short-distance structure of field correlations,
without first reconstructing the metric and applying
the usual curvature formulas. Such an identification
would make explicit how correlations encode this tensor
and provide a concrete starting point for investigating
descriptions of gravity in which correlations are treated
as primary variables. 
}

\tbb{
The standard metric-recovery prescription does not
immediately provide this identification. It uses mixed
second derivatives, with one derivative acting at each
point, which are tensorial without introducing a
connection. However, its coincidence limit isolates the
leading singular behaviour and discards the subleading
information. Taking higher partial derivatives does not,
in general, preserve this tensorial character, whereas
replacing them with covariant derivatives requires
knowledge of the connection. The challenge is therefore
to retain the subleading geometric information while
preserving a prescription that does not require a
previously reconstructed connection.
}

To overcome this limitation, we introduce a one-parameter covariant
regularization of the local Hadamard structure, controlled by an auxiliary UV
regulator $\ell_{0}$. Similar constructions have appeared in studies involving a fundamental or zero-point length
\cite{DeWitt81,Padmanabhan_1997,Kothawala_2013,Stargen_2015}; here, however, $\ell_{0}$ is
introduced only as a regulator, without modifying the background geometry or
assigning it the meaning of a physical minimal length. For finite $\ell_{0}$,
the resulting biscalar $W_{\ell_{0}}$ need not be the two-point function of a
Klein--Gordon field. Its purpose is instead to retain the subleading UV
information that is lost in the standard metric-recovery limit.

We show that the regulator dependence of $W_{\ell_{0}}$ provides direct access
to geometric information beyond the metric. In particular, for a massless
minimally coupled scalar field, the leading and first subleading terms recover
the metric and Einstein tensor, respectively. More generally, the Ricci
sector, as well as the connection and Riemann tensor, can be reconstructed
from coincidence derivatives of the regulated correlator. We also investigate
the matter side of semiclassical gravity and show how the renormalized
stress-energy tensor can be recovered from a regulated point-splitting
functional constructed from $W_{\ell_{0}}$, up to the usual finite
renormalization freedom
\cite{Wald1977BackReaction,Wald1978TraceAnomaly,Hollands_Wald2001, bibleHadamard}. These results show that the geometric and matter quantities entering the
semiclassical Einstein equations can be organized in terms of the same
regulated correlation structure.

This paper is organized as follows. In Sec.~\ref{sec:metric_recovery}, we
review the reconstruction of the spacetime metric from the short-distance
structure of the Wightman function and discuss its limitations. In
Sec.~\ref{sec:uv_deformation}, we introduce the covariant UV regularization used
throughout this work. In Sec.~\ref{sec:main}, we derive the reconstruction of
the Einstein and Ricci tensors from the regulator
dependence of $W_{\ell_{0}}$. In Sec.~\ref{sec:semiclassical}, we investigate
the corresponding formulation of the matter source in semiclassical
gravity. Our conclusions and outlook are presented in
Sec.~\ref{sec:conclusions}.

{\it Conventions}. For a biscalar $f(\mf x,\mf x')$, we use the shorthand
\begin{equation}
    f_{;\mu} \equiv \nabla_{\mu}f,
    \qquad
    f_{;\mu'} \equiv \nabla_{\mu'}f,
    \qquad
    f_{;\mu\nu'} \equiv \nabla_{\mu}\nabla_{\nu'}f,
\end{equation}
where $\nabla_{\mu}$ denotes the covariant derivative associated with
the Levi-Civita connection of the background metric $g_{\mu\nu}$.
Unprimed indices refer to derivatives with respect to the first
spacetime point, $\mf x$, while primed indices refer to derivatives
with respect to the second point, $\mf x'$. For ordinary partial
derivatives, we use
\begin{equation}
    f_{\mu} \equiv \partial_{\mu}f,
    \qquad
    f_{\mu'} \equiv \partial_{\mu'}f,
    \qquad
    f_{\mu\nu'} \equiv \partial_{\mu}\partial_{\nu'}f.
\end{equation}
\tbb{For a biscalar, first covariant derivatives and mixed second
derivatives acting at different points coincide with partial
derivatives:}
\begin{equation}
    \tbb{
    f_{;\mu}=f_{\mu},
    \qquad
    f_{;\mu'}=f_{\mu'},
    \qquad
    f_{;\mu\nu'}=f_{\mu\nu'}.
    }
\end{equation}
\tbb{By contrast, repeated differentiation at the same point generally
introduces connection terms. For example,}
\begin{equation}
    \tbb{
    \nabla_{\mu}\nabla_{\nu}f
    =
    \partial_{\mu}\partial_{\nu}f
    -
    \Gamma^{\rho}{}_{\mu\nu}\partial_{\rho}f.
    }
\end{equation}
\tbb{This distinction matters for the higher derivatives in this work.
We therefore retain both notations, using partial derivatives in the
reconstruction formulas to indicate when no independent knowledge
of the connection is required.}

The coincidence limit is denoted by
\begin{equation}
    [f] \equiv \lim_{\mf x'\to\mf x} f(\mf x,\mf x').
\end{equation}
We work with metric signature $(-,+,+,+)$.

\section{Metric recovery from two-point functions}
\label{sec:metric_recovery}

In this section, we review how the local geometry of spacetime is encoded in
the short-distance structure of quantum field correlations, as first established by Ref.~\cite{Saravani_2016} (see also \cite{Perche_Edu_geo_2022}). We begin with the Hadamard form of the Wightman function and recall how its leading ultraviolet
singularity determines Synge's world function and, consequently, the
spacetime metric. We then discuss the extent to which this construction can
be used to express the Einstein tensor in terms of field correlations. This
will identify the limitation of the standard metric-recovery formula and
motivate the ultraviolet regularization introduced in the next section.

On a spacetime $(\mathcal{M}, g)$, consider a real scalar quantum field $\hat{\phi}$ satisfying the
Klein--Gordon equation
\begin{equation}
    \bigl(\nabla^{\mu}\nabla_{\mu} - m^2 - \xi R\bigr)\hat{\phi}(\mf x) = 0,
\end{equation}
where $m$ is the field mass and $\xi$ is its coupling to the scalar
curvature. 

A state $\omega$ of the quantum field is a
normalized positive linear functional on the $\ast$-algebra
generated by the smeared fields $\hat{\phi}(f)$, with
$f \in C_0^\infty(\mathcal{M})$~\cite{Khavkine_2015}.
The smeared fields are formally written as
\begin{equation}
    \hat{\phi}(f)
    =
    \int_{\mathcal{M}} \dd V_{\mf x}\,
    f(\mf x)\hat{\phi}(\mf x),
\end{equation}
where $\dd V$ is the invariant spacetime volume element.
The two-point (or Wightman) distribution of the state is
defined by
\begin{equation}
    W(f,g)
    =
    \omega\bigl(\hat{\phi}(f)\hat{\phi}(g)\bigr),
    \qquad f,g \in C_0^\infty(\mathcal{M}).
\end{equation}
Its distributional kernel $W(\mf x,\mf x')$ satisfies
\begin{equation}
    W(f,g)
    =
    \int_{\mathcal{M}\times\mathcal{M}}
    \dd V_{\mf x}\,\dd V_{\mf x'}\,
    f(\mf x)g(\mf x')W(\mf x,\mf x'),
\end{equation}
where the integral denotes a distributional pairing.
Throughout this work, we write
\begin{equation}
    W(\mf x,\mf x')
    =
    \omega\bigl(\hat{\phi}(\mf x)\hat{\phi}(\mf x')\bigr)
\end{equation}
as shorthand for this distributional kernel and refer to
it simply as the two-point (or Wightman) function.

On a generic curved spacetime, the Hadamard condition provides a local criterion for selecting physically admissible states through the short-distance structure of their Wightman functions~\cite{fullingHadamard,fullingHadamard2,fewsterNecessityHadamard}. Motivated by the equivalence principle, it requires a universal local singularity structure whose leading term agrees, in local inertial coordinates, with that of the Minkowski vacuum.

This condition also provides the UV regularity
needed for the standard definition of local composite observables.
For fixed geometry and field parameters, the two-point functions of
any two Hadamard states differ by a smooth biscalar. Their common
singular part can therefore be subtracted using a state-independent
prescription, yielding finite renormalized expectation values of
observables such as the stress-energy tensor, up to the usual finite
local renormalization freedom~\cite{bibleHadamard}. This is particularly
relevant when the field is used as a source in semiclassical gravity.

More precisely, let $\mf x$ and $\mf x'$
lie in the same convex normal neighbourhood, so that they are connected by
a unique geodesic \tbb{contained in that neighbourhood}
and Synge's world function
$\sigma(\mf x,\mf x')$ is well defined~\cite{poissonbible}\tbb{.
Our coincidence limit calculations presented below require only this local definition.
For a rigorous treatment of its extension to larger domains, see
Ref.~\cite{Moretti2021GlobalHadamard}.}\footnote{The Hadamard condition has
an analogue in arbitrary spacetime dimension. We suppress the usual
$\ii\epsilon$ prescription because only local coincidence-limit
expressions will be considered below.}
\begin{align}
    W(\mf x,\mf x')
    &=
    \frac{1}{8\pi^2}
    \Biggl[
        \frac{\Delta^{1/2}(\mf x,\mf x')}
        {\sigma(\mf x,\mf x')}
        +V(\mf x,\mf x')
        \log\!\left(
            \frac{\sigma(\mf x,\mf x')}{L_{\mathrm{H}}^2}
        \right)
        \nonumber\\
    &\hspace{3.4cm}
        +w(\mf x,\mf x')
    \Biggr],
    \label{eq:hadamard_first_time}
\end{align}
where $\Delta(\mf x,\mf x')$ is the van Vleck determinant and $L_{\mathrm{H}}$ is
an arbitrary reference length. The biscalars $V$ and $w$ admit regular
expansions in powers of Synge's world function,
\begin{equation}
    V(\mf x,\mf x')
    =
    \sum_{n=0}^{\infty}
    V_n(\mf x,\mf x')\sigma(\mf x,\mf x')^n,
    \label{eq:expansion_V}
\end{equation}
and
\begin{equation}
    w(\mf x,\mf x')
    =
    \sum_{n=0}^{\infty}
    w_n(\mf x,\mf x')\sigma(\mf x,\mf x')^n.
\end{equation}
The biscalar $V$ is determined locally by the field equation and the
background geometry, whereas $w$ contains the dependence on the quantum
state.

Knowledge of Synge's world function is sufficient to reconstruct the local
geometry of spacetime. Indeed, one has~\cite{poissonbible}
\begin{equation}
    g_{\mu\nu}(\mf x)
    =
    -[\sigma_{\mu\nu'}(\mf x,\mf x')].
    \label{eq:metric_from_sigma}
\end{equation}
Since the leading Hadamard singularity is governed by
$\sigma(\mf x,\mf x')$, it follows that the metric can be reconstructed
directly from the ultraviolet behaviour of the two-point function. This observation was developed for scalar-field propagators in
Ref.~\cite{Saravani_2016}, and later extended to the Wightman function by
Ref.~\cite{Perche_Edu_geo_2022}. In particular, for a scalar field in a
Hadamard state in $D$ spacetime dimensions, one has
\begin{equation}
    g_{\mu\nu}
    =
    -\frac{1}{2}
    \left(
        \frac{
            \Gamma\!\left(\frac{D}{2}-1\right)
        }{
            4\pi^{D/2}
        }
    \right)^{\frac{2}{D-2}}
    \left[
        \partial_{\mu}\partial_{\nu'}
        \bigl(W(\mf x,\mf x')\bigr)^{\frac{2}{2-D}}
    \right].
    \label{eq:metric_from_W}
\end{equation}
Thus, the local notion of spacetime distance can be inferred from the
strength of field correlations. This result supports the broader viewpoint
that the metric need not be regarded as a primitive structure, but may be
encoded in the correlation functions of quantum fields
\cite{Kempf_2021}.

These results motivate the following question: can Einstein's equations be
written directly in terms of field correlations? As a first step, we focus on
their geometric side, namely, on the Einstein tensor.

In principle, Eq.~\eqref{eq:metric_from_W} already provides all the
information needed to reconstruct $G_{\mu\nu}$. Once the metric has been
obtained on an open neighbourhood, one may calculate the connection, then
the Riemann tensor, and finally form the Einstein tensor. However, this
procedure is indirect. When written entirely in terms of the two-point
function, it would involve several derivatives of coincidence limits and
would lead to a highly cumbersome expression.

More importantly, the metric-recovery formula uses only the leading
coincidence behaviour of the Wightman function. At this order, the Hadamard
singularity is controlled by $\sigma(\mf x,\mf x')$ and therefore contains
precisely the information required to recover $g_{\mu\nu}$. It does not,
however, isolate higher-order curvature quantities such as
$R_{\mu\nu}$ or $G_{\mu\nu}$. The curvature dependence of the Hadamard form
appears only at subleading orders in the short-distance expansion.

The preceding limitation can be made precise for a natural class of direct
constructions. In four spacetime dimensions, suppose that, for fixed field
parameters $m$ and $\xi$, there exists a universal $C^{2}$ pointwise
function $\Phi_{m,\xi}$ such that
\begin{equation}
    G_{\mu\nu}(\mf x)
    =
    \left[
        \partial_\mu\partial_{\nu'}
        \Phi_{m,\xi}\bigl(W(\mf x,\mf x')\bigr)
    \right].
    \label{eq:direct_Einstein_ansatz}
\end{equation}
Here, universal means that $\Phi_{m,\xi}$ may depend on the fixed field
parameters, but not on the spacetime geometry, the
quantum state, the points $\mf x,\mf x'$, or any auxiliary regulator. \tbb{This ansatz has the same mathematical structure as the
metric recovery formula, but targets the Einstein tensor instead.
Specifically, in four spacetime dimensions,
Eq.~\eqref{eq:metric_from_W} reads}
\begin{equation}
    \tbb{
    g_{\mu\nu}
    =
    \left[
        \partial_\mu\partial_{\nu'}
        \Phi_{\mathrm{met}}\bigl(W(\mf x,\mf x')\bigr)
    \right],
    \qquad
    \Phi_{\mathrm{met}}(z)
    =
    -\frac{1}{8\pi^{2}z}.
    }
\end{equation}
\tbb{The transformation $\Phi_{\mathrm{met}}$ is independent of
$m$ and $\xi$, since the leading Hadamard singularity underlying
metric recovery is universal. The question posed by
Eq.~\eqref{eq:direct_Einstein_ansatz} is whether a different universal
pointwise transformation can recover $G_{\mu\nu}$ through the same
mixed second-derivative construction.}
In Appendix~\ref{sec:proof_no_go}, we prove that no universal function
$\Phi_{m,\xi}$ can satisfy Eq.~\eqref{eq:direct_Einstein_ansatz}. Thus, a
single scalar reparametrization of the Wightman function cannot
directly isolate the Einstein tensor through the coincidence limit of derivatives.

This observation motivates a different strategy. Rather than seeking a
single scalar reparametrization of the Wightman function, we
introduce an auxiliary covariant ultraviolet regulator $\ell_{0}$ and study
the resulting one-parameter family of deformed correlation functions. The leading regulator
dependence recovers the metric, while the first subleading dependence
isolates geometric information beyond the metric. We will use this
hierarchy to identify the Ricci and Einstein tensors directly within the
regulated short-distance structure of field correlations.

\section{A covariant UV regularization of the Hadamard form}
\label{sec:uv_deformation}

The discussion of the previous section motivates the introduction of a
controlled regularization of the UV structure of the Wightman
function. Our goal is to access the subleading short-distance behaviour of
the correlator, which is not captured by the leading coincidence limit used
to reconstruct the metric. To this end, we introduce an auxiliary parameter
$\ell_{0}>0$, with dimensions of length, which acts as an ultraviolet
regulator and will be removed at the end of the calculation. 

We implement this regularization by replacing Synge's world function in the
Hadamard structure by the biscalar
\begin{equation}
    \mathcal{L}_{\ell_{0}}(\mf x, \mf x')
    =
    \sigma(\mf x,\mf x')+\frac{\ell_{0}^2}{2},
    \label{eq:effective_distance}
\end{equation}
where $\mathcal{L}_{\ell_{0}}(\mf x,\mf x')$ is simply a convenient notation.

The algebraic shift in Eq.~\eqref{eq:effective_distance} is formally
similar to constructions considered in
Refs.~\cite{DeWitt81,Kothawala_2013,Stargen_2015}. In those approaches,
$\ell_{0}$ is interpreted as a minimal, or zero-point, spacetime length and
is used to construct a modified effective geometry. Our use of the same
shift is deliberately different: $\ell_{0}$ is only an auxiliary UV
regulator, while the original background is kept fixed.

This distinction is central to our construction. No physical interpretation is
assigned to finite $\ell_{0}$: it is neither a fundamental length, a finite
experimental resolution, nor a modification of the background geometry.
Rather, $\ell_{0}$ is an auxiliary mathematical parameter, introduced in the
same methodological spirit as the regulator in dimensional regularization (see, e.g.,~\cite{Bollini1972,tHooftVeltman1972}).
It labels an intermediate family of expressions that organizes the
ultraviolet expansion and is removed after the relevant coefficients have
been isolated.

We define the UV-regulated Wightman function by
\begin{align}
 W_{\ell_{0}}(\mf x,\mf x')
    & =
    \frac{1}{8\pi^2}
    \Biggl[
        \frac{\Delta^{1/2}(\mf x,\mf x')}
        {\mathcal{L}_{\ell_{0}}(\mf x,\mf x')}
        +V(\mf x,\mf x')
        \log\!\left(
        \frac{\mathcal{L}_{\ell_{0}}(\mf x,\mf x')}{L_{\mathrm{H}}^2}
        \right)
        \nonumber \\
        &\hspace{3.5cm}
        +w(\mf x,\mf x')
    \Biggr].
    \label{eq:hadamard_regulated}
\end{align}
The reference scale $L_{\mathrm{H}}$ appearing in the logarithm is the same arbitrary
Hadamard scale introduced previously in Eq.~\eqref{eq:hadamard_first_time} and should not be confused with the UV
regulator $\ell_{0}$.

From now on, we shall refer to $W_{\ell_{0}}$ as the UV-regulated Wightman function. Nonetheless, it is important to stress that, for finite $\ell_{0}$, it need not be a genuine
two-point function of the Klein--Gordon field. Rather,
$W_{\ell_{0}}(\mf x,\mf x')$ is an auxiliary biscalar obtained by deforming the
ultraviolet Hadamard structure of $W(\mf x,\mf x')$.
With this interpretation in mind, a few observations are in order. First,
for every finite $\ell_{0}>0$, $W_{\ell_{0}}$ is finite in the coincidence
limit. Indeed, since
\begin{equation}
    \mathcal{L}_{\ell_{0}}(\mf x,\mf x)
    =
    \frac{\ell_{0}^2}{2},
\end{equation}
it follows that
\begin{align}
    [W_{\ell_{0}}]
    &=
    \frac{1}{8\pi^2}
    \left(
        \frac{2}{\ell_{0}^2}
        +
        [V]\log\!\left(\frac{\ell_{0}^2}{2L_{\mathrm{H}}^2}\right)
        +
        [w]
    \right).
    \label{eq:regulated_coincidence_limit}
\end{align}
Thus, the diagonal coincidence singularity is replaced by a finite,
regulator-dependent expression. However, globally, the singularity is
displaced from the background light cone
$\sigma(\mf x,\mf x')=0$ to the locus
\begin{equation}
    \mathcal{L}_{\ell_{0}}(\mf x,\mf x')=0
    \implies
    \sigma(\mf x,\mf x') = -\frac{\ell_{0}^2}{2}.
\end{equation}
In Minkowski spacetime,
\begin{equation}
    \sigma_0(\mf x,\mf x')
    =
    \frac{-(\Delta t)^2+|\Delta\boldsymbol{x}|^2}{2},
    \label{eq:flat_Synge}
\end{equation}
and the displaced singular locus becomes
\begin{equation}
    (\Delta t)^2-|\Delta\boldsymbol{x}|^2=\ell_{0}^2.
    \label{eq:shifted_singularity_surface}
\end{equation}
Thus, in this case, the light-cone singularity is shifted to a timelike
hyperboloid. This feature is formally reminiscent of the displacement of
light-cone singularities discussed by DeWitt~\cite{DeWitt81}. The
regularization should therefore be understood as a prescription that regulates
coincidence limits, rather than as a globally regular two-point function.

The role of $\ell_{0}$ as an ultraviolet regulator can also be illustrated in
a simple flat-spacetime example. Consider a massless scalar field in
$3+1$-dimensional Minkowski spacetime. In inertial coordinates
$\mf x=(t,\boldsymbol{x})$, the standard Wightman function is
\begin{equation}
    W_0(\mf x,\mf x')
    =
    \frac{1}{8\pi^2}
    \frac{1}{\sigma_0(\mf x,\mf x')},
\end{equation}
where $\sigma_0$ is the flat Synge's world function given in
Eq.~\eqref{eq:flat_Synge}. The UV-regulated prescription amounts to
replacing $\sigma_0$ by $\mathcal{L}_{\ell_{0}}$:
\begin{equation}
    W_{\ell_{0}}(\mf x,\mf x')
    \equiv
    \frac{1}{8\pi^2}
    \frac{1}{\mathcal{L}_{\ell_{0}}(\mf x,\mf x')}.
    \label{eq:Wightmanl0mink}
\end{equation}

Let us examine how this replacement modifies the two-point function in
momentum space. Consider the Fourier transform
\begin{equation}
    \widetilde{f}(\mf p)
    =
    \int \dd^4 \mf x\,
    e^{\ii \mf p\cdot \mf x}
    f(\mf x),
\end{equation}
where $\mf p\cdot\mf x\equiv \eta_{\mu\nu}p^\mu x^\nu$. Applying this
convention to Eq.~\eqref{eq:Wightmanl0mink}, one obtains
\cite{Padmanabhan_1997}
\begin{align}
    \widetilde{W}_{\ell_{0}}(\mf p)
   &  =
    2\pi\,\Theta(-p^0)\,\delta(\mf p^2)
    \nonumber \\ & \quad -
    \pi \ell_{0}\,
    \frac{
        J_1\!\left(\ell_{0}\sqrt{-\mf p^2}\right)
    }{
        \sqrt{-\mf p^2}
    }\,
    \Theta(-p^0-|\boldsymbol{p}|),
    \label{eq:fourier_regulated_wightman}
\end{align}
where $\mf p^2=\eta_{\mu\nu}p^\mu p^\nu$, $\Theta$ is the Heaviside step
function, and $J_\alpha$ is the Bessel function of the first kind. A
derivation is given in Appendix~\ref{sec:fourier_stuff}.

The first term in Eq.~\eqref{eq:fourier_regulated_wightman} is the
\tbb{standard} Fourier transform of the
\tbb{massless vacuum} Wightman function (\tbb{see, e.g.,~\cite{PeskinSchroeder1995}}).
The second term
contains the correction induced by the UV regularization:
\begin{equation}
    \widetilde{\Delta W}_{\ell_{0}}(\mf p)
    =
    -
    \pi \ell_{0}\,
    \frac{
        J_1\!\left(\ell_{0}\sqrt{-\mf p^2}\right)
    }{
        \sqrt{-\mf p^2}
    }\,
    \Theta(-p^0-|\boldsymbol{p}|).
    \label{eq:fourier_main_text}
\end{equation}
This correction has support inside the negative-frequency timelike region,
$p^0<-|\boldsymbol{p}|$, and is controlled by the invariant momentum scale  $q\equiv \sqrt{-\mf p^2}$. In the UV regime $q\to\infty$, the Bessel function satisfies
\begin{equation}
    J_1(\ell_{0} q)
    \sim
    \sqrt{\frac{2}{\pi \ell_{0} q}}
    \cos\!\left(\ell_{0} q-\frac{3\pi}{4}\right),
\end{equation}
and therefore
\begin{equation}
    \widetilde{\Delta W}_{\ell_{0}}(\mf p)
    \sim
    -
    \sqrt{2\pi\ell_{0}}\,
    \frac{
        \cos\!\left(\ell_{0} q-\frac{3\pi}{4}\right)
    }{
        q^{3/2}
    }.
    \label{eq:fourier_uv_asymptotic}
\end{equation}
Thus, the finite-$\ell_{0}$ regularization induces an oscillatory correction
whose amplitude is suppressed as $q^{-3/2}$ at large invariant momentum.
Together with the finite coincidence limit in
Eq.~\eqref{eq:regulated_coincidence_limit}, this illustrates the role of
$\ell_{0}$ as a covariant UV regulator.

The affine prescription in Eq.~\eqref{eq:effective_distance} is the
specific regularization used in the next section. It is useful, however, to state the
construction for a more general family of functions $f_{\ell_{0}}(\sigma)$.
For every fixed $\ell_{0}$, we require $f_{\ell_{0}}(0)>0$ and require
$f_{\ell_{0}}(\sigma)$ to be nonvanishing in a sufficiently small
neighbourhood of $\sigma=0$. We further require that
$f_{\ell_{0}}(\sigma)\to\sigma$ as $\ell_{0}\to0$, so that the original
Hadamard form is recovered. Writing $f$ for $f_{\ell_{0}}$, the corresponding
UV-regulated two-point function is
\begin{align}
W_{f}(\mf x,\mf x')
&=
\frac{1}{8\pi^2}
\Biggl[
    \frac{\Delta^{1/2}(\mf x,\mf x')}
    {f\bigl(\sigma(\mf x,\mf x')\bigr)}
    \nonumber\\
&\qquad
    +V(\mf x,\mf x')
    \log\!\left(
        \frac{f\bigl(\sigma(\mf x,\mf x')\bigr)}{L_{\mathrm{H}}^2}
    \right)
    +w(\mf x,\mf x')
\Biggr],
\label{eq:hadamard_regulated_general_main}
\end{align}
The next section shows how
\tbb{geometric information beyond the metric can be
extracted from the coincidence derivatives of this family of
UV-regulated two-point functions by examining their asymptotic
behaviour as $\ell_0\to0$.}

\section{The Einstein tensor from the UV-regulated Wightman function}
\label{sec:main}

We now examine the regulator dependence of the coincidence derivatives of
the UV-regulated two-point function. The relevant calculation is presented in
Appendix~\ref{sec:calculations}, where the general expression of Eq.~\eqref{eq:hadamard_regulated_general_main} is
considered. For the concrete choice of UV regularization introduced in Eq.~\eqref{eq:hadamard_regulated}, one finds
\begin{equation}
    2\pi^2 \ell_{0}^4
    \bigl[
        \partial_{\mu}\partial_{\nu'}W_{\ell_{0}}
    \bigr]
    =
    g_{\mu\nu}
    -
    \frac{\ell_{0}^2}{12}
    \left(
        R_{\mu\nu}
        +
        6[V]g_{\mu\nu}
    \right)
    +
    \ell_{0}^4Y^{(\ell_{0})}_{\mu\nu},
    \label{eq:effective_metric_expansion}
\end{equation}
where
\begin{equation}
    [V]
    =
    \frac{m^2}{2}
    +
    \frac{R}{2}
    \left(
        \xi-\frac{1}{6}
    \right),
    \label{eq:V_m_xi}
\end{equation}
and
\begin{equation}
    Y^{(\ell_{0})}_{\mu\nu}
    =
    \frac{1}{4}
    [V_{;\mu\nu'}]
    \log\!\left(
        \frac{\ell_{0}^2}{2L_{\mathrm{H}}^2}
    \right)
    +
    \frac{1}{4}[w_{;\mu\nu'}].
    \label{eq:Y_l0_definition}
\end{equation}
\tbb{Here we have used
$\nabla_{\mu}\nabla_{\nu'}W_{\ell_{0}}
=\partial_{\mu}\partial_{\nu'}W_{\ell_{0}}$, since the two derivatives
act at different points on a biscalar}.
The tensor $[V_{\mu\nu'}]$ is a local expression involving higher-order
curvature terms, whose explicit form is given in
Appendix~\ref{sec:calculations}. The term $[w_{\mu\nu'}]$, on the other
hand, contains the state dependence of the two-point function.

Equation~\eqref{eq:effective_metric_expansion} displays a hierarchy of
geometric information in the regulator dependence of
$W_{\ell_{0}}(\mf x,\mf x')$. The leading term determines the metric, whereas
the first subleading term contains the Ricci tensor. Since
$Y^{(\ell_{0})}_{\mu\nu}$ grows at most logarithmically as $\ell_{0}\to0$, the
last term in Eq.~\eqref{eq:effective_metric_expansion} does not contribute
to the first subleading coefficient, which is therefore state-independent.

The leading term in Eq.~\eqref{eq:effective_metric_expansion} gives
\begin{equation}
    g_{\mu \nu}
    =
    \lim_{\ell_{0} \to 0}
    2\pi^2 \ell_{0}^{4}
    [\partial_{\mu}\partial_{\nu'}W_{\ell_{0}}(\mf x,\mf x')].
    \label{eq:metric_from_W_l0}
\end{equation}

We now isolate the first subleading coefficient in
Eq.~\eqref{eq:effective_metric_expansion}. Define
\begin{equation}
    C_{\mu\nu}(m,\xi)
    \equiv
    -12
    \lim_{\ell_{0}\to0}
    \frac{
        2\pi^2\ell_{0}^4
        [\partial_\mu\partial_{\nu'}W_{\ell_{0}}]
        -g_{\mu\nu}
    }{\ell_{0}^2}.
    \label{eq:C_definition}
\end{equation}
The metric in this expression is determined by
Eq.~\eqref{eq:metric_from_W_l0}. Moreover, the contribution from
$Y^{(\ell_{0})}_{\mu\nu}$ vanishes in the limit defining
$C_{\mu\nu}(m,\xi)$. Using Eq.~\eqref{eq:V_m_xi}, we obtain
\begin{equation}
    C_{\mu \nu}(m,\xi)
    =
    R_{\mu\nu}
    +
    \left[
        3m^2
        +
        3R\left(\xi-\frac{1}{6}\right)
    \right]g_{\mu\nu}.
    \label{eq:C_general}
\end{equation}

The geometric quantity isolated by $C_{\mu\nu}(m,\xi)$ depends on the field
parameters. For a massless conformally coupled scalar field,
\begin{equation}
    C_{\mu\nu}(0,1/6)
    =
    R_{\mu\nu}.
    \label{eq:C_Ricci}
\end{equation}
For a massless minimally coupled scalar field, one instead finds
\begin{equation}
    C_{\mu\nu}(0,0)
    =
    R_{\mu\nu}
    -
    \frac{1}{2}Rg_{\mu\nu}
    =
    G_{\mu\nu}.
    \label{eq:C_Einstein}
\end{equation}
Thus, for this choice of field parameters, the first subleading regulator
dependence of the mixed second derivative of the UV-regulated Wightman function directly
yields the Einstein tensor.

Writing Eq.~\eqref{eq:C_Einstein} explicitly in terms of $W_{\ell_{0}}$ gives
\begin{align}
    G_{\mu\nu}
    &=
    -12
    \lim_{\ell_{0}\to 0}
    \Biggl(
        2\pi^2\ell_{0}^{2}
        [\partial_{\mu}\partial_{\nu'}W_{\ell_{0}}(\mf x,\mf x')]
        \nonumber\\
    &\hspace{1.6cm}
        -
        \frac{
            \displaystyle
            \lim_{\lambda\to 0}
            2\pi^2\lambda^{4}
            [\partial_{\mu}\partial_{\nu'}W_{\lambda}(\mf x,\mf x')]
        }{
            \ell_{0}^{2}
        }
    \Biggr).
    \label{eq:could_be_einstein}
\end{align}
The inner limit is precisely the metric-reconstruction formula,
Eq.~\eqref{eq:metric_from_W_l0}. Therefore,
Eq.~\eqref{eq:could_be_einstein} contains no independent geometric input
beyond the one-parameter family of UV-regulated Wightman functions.

Equation~\eqref{eq:could_be_einstein} is the main result of this section.
It shows that the Einstein tensor of the background spacetime can be
obtained directly from the regulator dependence of the two-point function of
a massless minimally coupled scalar field. 

It is worth emphasizing that the choice $m=\xi=0$ is not essential to the
geometric reconstruction developed above. It is distinguished only because
it yields the Einstein tensor directly from the first subleading regulator
coefficient. For generic values of $m$ and $\xi$, the reconstruction of
$G_{\mu\nu}$ additionally requires contractions with the inverse metric.
This does not introduce independent geometric input, since the inverse metric
can itself be recovered from the UV-regulated Wightman function.

To see this, consider the coincidence limit of the mixed derivatives of
$W_{\ell_{0}}$, which defines the rank-two tensor
\begin{equation}
    \bigl[\partial\partial'W_{\ell_{0}}\bigr]_{\alpha\beta}
    \equiv
    \lim_{\mf x'\to\mf x}
    \partial_\alpha\partial_{\beta'}
    W_{\ell_{0}}(\mf x,\mf x').
\end{equation}
\tbb{To establish its invertibility, note that
Eq.~\eqref{eq:effective_metric_expansion} implies}
\begin{equation}
    \tbb{
    2\pi^2\ell_0^4
    \bigl[\partial\partial'W_{\ell_0}\bigr]_{\alpha\beta}
    \longrightarrow g_{\alpha\beta}
    \qquad (\ell_0\to0).
    }
\end{equation}
\tbb{Since the metric is nondegenerate, continuity of the determinant
ensures that, at each spacetime point, the normalized tensor is
invertible for sufficiently small $\ell_0>0$. The same therefore
holds for $\bigl[\partial\partial'W_{\ell_0}\bigr]_{\alpha\beta}$,
whose inverse is defined by}
\begin{equation}
    \left(
        \bigl[\partial\partial'W_{\ell_0}\bigr]^{-1}
    \right)^{\mu\alpha}
    \bigl[\partial\partial'W_{\ell_0}\bigr]_{\alpha\nu}
    =
    \delta^\mu{}_\nu.
    \label{eq:def_inverse}
\end{equation}
\tbb{Continuity of matrix inversion then implies that the inverse
of the normalized tensor converges to $g^{\mu\nu}$. Thus,}
\begin{equation}
    g^{\mu\nu}
    =
    \lim_{\ell_{0}\to0}
    \frac{1}{2\pi^2 \ell_{0}^4}
    \left(
        \bigl[\partial\partial'W_{\ell_{0}}\bigr]^{-1}
    \right)^{\mu\nu}.
    \label{eq:inverse_metric_reconstruction}
\end{equation}

\tbb{The metric determinant can be reconstructed similarly. In four
spacetime dimensions, continuity of the determinant gives}
\tbb{\begin{equation}
    g \equiv \det(g_{\mu\nu})
    =
    \lim_{\ell_0\to0}
    \left(2\pi^2\ell_0^4\right)^4
    \det\!\left(
        \bigl[\partial\partial'W_{\ell_0}\bigr]_{\alpha\beta}
    \right).
    \label{eq:metric_determinant_reconstruction}
\end{equation}}
\tbb{Since $g<0$ for a Lorentzian metric, the determinant on the
right-hand side is also negative for sufficiently small $\ell_0$.
The volume density therefore follows as}
\tbb{
\begin{equation}
    \sqrt{-g}
    =
    \lim_{\ell_0\to0}
    \left(2\pi^2\ell_0^4\right)^2
    \sqrt{
        -\det\!\left(
            \bigl[\partial\partial'W_{\ell_0}\bigr]_{\alpha\beta}
        \right)
    }.
    \label{eq:volume_density_reconstruction}
\end{equation}
}
We may therefore take the trace of Eq.~\eqref{eq:C_definition} without
introducing any additional geometric data. Defining
\begin{equation}
    C(m,\xi)
    \equiv
    g^{\mu\nu}C_{\mu\nu}(m,\xi),
\end{equation}
we obtain
\begin{equation}
    C(m,\xi)
    =
    (12\xi-1)R+12m^2.
\end{equation}
Hence, whenever $\xi\neq1/12$, the scalar curvature is recovered as
\begin{equation}
    R
    =
    \frac{C(m,\xi)-12m^2}{12\xi-1}.
    \label{eq:R_recover}
\end{equation}
Substituting this expression into Eq.~\eqref{eq:C_general} yields
\begin{align}
    R_{\mu\nu}
    &=
    C_{\mu\nu}(m,\xi)
    \nonumber\\
    &\quad
    -
    \left[
        3m^2
        +
        3\left(\xi-\frac{1}{6}\right)
        \frac{C(m,\xi)-12m^2}{12\xi-1}
    \right]g_{\mu\nu}.
    \label{eq:ricci_recover}
\end{align}
Consequently, the Einstein tensor can be written as
\begin{equation}
    G_{\mu\nu}
    =
    C_{\mu\nu}(m,\xi)
    -
    \left[
        3m^2
        +
        3\xi
        \frac{C(m,\xi)-12m^2}{12\xi-1}
    \right]g_{\mu\nu}.
\end{equation}
\tbb{
Substituting Eq.~\eqref{eq:C_definition} and its trace gives
the corresponding reconstruction formula for arbitrary $m$
and $\xi\neq1/12$:}
\begin{align}
  G_{\mu\nu}
    &=
    -12\lim_{\ell_0\to0}
    \Biggl[
        2\pi^2\ell_0^2
        \Biggl(
            [\partial_\mu\partial_{\nu'}W_{\ell_0}]
    \nonumber\\
    &\hspace{2.3cm}
            -
            \frac{3\xi}{12\xi-1}
            g_{\mu\nu}g^{\alpha\beta}
            [\partial_\alpha\partial_{\beta'}W_{\ell_0}]
        \Biggr)
    \nonumber\\
    &\hspace{1.6cm}
        +
        \frac{g_{\mu\nu}}{(12\xi-1)\ell_0^2}
    \Biggr]
    +
    \frac{3m^2}{12\xi-1}g_{\mu\nu}.
    \label{eq:Einstein_from_regulated_W_general}
\end{align}
Here $W_{\ell_0}$ is the UV-regulated Wightman function for
the field with mass $m$ and curvature coupling $\xi$.
The metric and its inverse are determined by
Eqs.~\eqref{eq:metric_from_W_l0} and
\eqref{eq:inverse_metric_reconstruction}, respectively,
so this expression requires no independent geometric input.
For $m=\xi=0$, it reduces to
Eq.~\eqref{eq:could_be_einstein}.

Thus, apart from the degenerate value $\xi=1/12$, knowledge of the
one-parameter family $W_{\ell_0}$ is sufficient to determine
$R$, $R_{\mu\nu}$, and $G_{\mu\nu}$
\tbb{through $C_{\mu \nu}(m, \xi)$. At $\xi=1/12$, its trace is $C(m,1/12)=12m^2$ and therefore
does not determine $R$.}

In Appendix~\ref{sec:riemann}, we further show that the inverse-metric
reconstruction in Eq.~\eqref{eq:inverse_metric_reconstruction} can be used
together with higher coincidence derivatives of the UV-regulated Wightman
function to reconstruct the connection and the Riemann tensor.

\tbb{Therefore, the information encoded in $W_{\ell_{0}}$ suffices to reconstruct curvature tensors, including the Einstein tensor, without requiring the metric as an independent input. In particular, the Einstein tensor can be identified directly without computing it from spacetime derivatives of the reconstructed metric. This identification is particularly significant because of the tensor's role in Einstein's equations. Our result shows that it is directly encoded in the subleading singularity structure of the scalar field's two-point function, providing a concrete link between quantum field correlations and the geometry underlying classical gravity.}

\section{Semiclassical gravity}
\label{sec:semiclassical}

The preceding section shows that the geometric side of Einstein's
equations can be reconstructed from
\tbb{the coincidence derivatives of a family of
UV-regulated two-point functions as the regulator is removed}.
A natural next question is whether the source for gravity can be
expressed within the same framework. Semiclassical gravity
provides the standard effective setting in which a classical spacetime is
sourced by the renormalized expectation value of a quantum stress-energy
tensor.
\tbb{Quantum field theory on prescribed curved backgrounds predicts
cosmological particle creation and Hawking radiation, while
semiclassical gravity provides a framework for studying their
backreaction on spacetime geometry, including black-hole evaporation}
\cite{Parker1968,ParkerOG,Hawking1975,HawkingRadiation,Wald1977BackReaction,Wald1978TraceAnomaly}.
\tbb{These applications motivate the study of semiclassical gravity
but do not constitute experimental confirmation of its validity;
in particular, the evaporation of astrophysical black holes has not
been directly observed.}

\tbb{A theoretical motivation for semiclassical gravity comes from the large-$N$
expansion, in which gravity is coupled to $N$ independent, identical
free quantum fields. At leading order, with $N G$ held fixed,
where $G$ is Newton's gravitational constant, the geometry is
treated classically while the matter fields remain fully
quantized~(see, e.g,~\cite{HuVerdaguer2008}). This provides a formal basis for
the semiclassical approximation. Its validity nevertheless requires
metric fluctuations to remain sufficiently small on the scales
of interest~\cite{anderson2003,Huetal2004,HuVerdaguer2008}.}

It is therefore worthwhile to investigate to what extent the semiclassical
Einstein equation can itself be reformulated in terms of UV-regulated field
correlations.

For definiteness, we consider a real, massless, minimally coupled scalar
field in four spacetime dimensions. We work in natural units and omit a cosmological-constant term. The quantized stress-energy
tensor is
\begin{equation}
    \hat{T}_{\mu \nu}
    =
    \nabla_{\mu}\hat{\phi}\nabla_{\nu}\hat{\phi}
    -
    \frac{g_{\mu \nu}}{2}
    \nabla^{\alpha}\hat{\phi}\nabla_{\alpha}\hat{\phi}.
\end{equation}
Formally, its expectation value in a given field state can be written in
terms of the Wightman function as
\begin{equation}
    \langle \hat{T}_{\mu \nu}\rangle
    =
    \lim_{\mf x' \to \mf x}
    \left[
        \nabla_{\mu}\nabla_{\nu'}
        -
        \frac{g_{\mu \nu}}{2}
        \nabla^{\alpha}\nabla_{\alpha'}
    \right]
    W(\mf x,\mf x').
    \label{eq:Tmunu_bare}
\end{equation}
\tbb{Here we write the coincidence limit explicitly to distinguish it
from the square brackets delimiting the differential operator.}
Equation~\eqref{eq:Tmunu_bare} is a shorthand for the standard covariant,
symmetric point-splitting prescription, including the parallel transport of
primed indices before the coincidence limit is taken. As it stands, however,
the expression is divergent for every Hadamard state.

The Hadamard form in Eq.~\eqref{eq:hadamard_first_time} provides a local
renormalization prescription.
\tbb{In Hadamard point splitting, one first subtracts the universal
singular part of the two-point function, determined by the local
geometry and field parameters~\cite{Christensen1976,bibleHadamard}.
Writing this singular part as}
\begin{equation}
    \tbb{
    H_g(\mf x,\mf x')
    =
    \frac{1}{8\pi^2}
    \left[
        \frac{\Delta^{1/2}(\mf x,\mf x')}{\sigma(\mf x,\mf x')}
        +
        V(\mf x,\mf x')
        \log\!\left(
            \frac{\sigma(\mf x,\mf x')}{L_{\mathrm{H}}^2}
        \right)
    \right],
    }
\end{equation}
\tbb{with the same distributional prescription as in the Wightman
function, the difference is}
\begin{equation}
    \tbb{
    W(\mf x,\mf x')-H_g(\mf x,\mf x')
    =
    \frac{1}{8\pi^2}w(\mf x,\mf x').
    }
\end{equation}
\tbb{For a Hadamard state, this remainder is smooth. Applying the
point splitting differential operator in Eq.~\eqref{eq:Tmunu_bare} to it therefore yields a
finite coincidence limit. A fixed local geometric term must also
be included to ensure conservation, since the Hadamard parametrix $H_{g}$
is not generally an exact bisolution of the Klein-Gordon equation.}
In the prescription adopted here, the
renormalized expectation value is~\cite{bibleHadamard}
\begin{align}
    \langle \mathopen{:}\hat{T}_{\mu \nu}\mathclose{:}\rangle
    &=
    \frac{1}{8\pi^2}
    \lim_{\mf x' \to \mf x}
    \left[
        \nabla_{\mu}\nabla_{\nu'}
        -
        \frac{g_{\mu \nu}}{2}
        \nabla^{\alpha}\nabla_{\alpha'}
    \right]
    w(\mf x,\mf x')
    \nonumber \\
    &\quad
    +
    \frac{[V_1]}{4\pi^2}g_{\mu \nu}
    +
    \Theta^{(L_{\mathrm{H}})}_{\mu \nu}.
    \label{eq:renormalizedTmunu}
\end{align}
Here $w(\mf x,\mf x')$ is the smooth, state-dependent part of the Hadamard
form.
\tbb{Because the subtracted parametrix is not generally an exact
bisolution of the Klein--Gordon equation, the differentiated remainder
is not conserved by itself. For the massless, minimally coupled field
considered here, one can show that~
\cite{bibleHadamard}}
\begin{align}
    \tbb{
    \nabla^{\nu}
    \left\{
        \frac{1}{8\pi^2}
        \lim_{\mf x'\to\mf x}
        \left[
            \nabla_{\mu}\nabla_{\nu'}
            -
            \frac{g_{\mu\nu}}{2}
            \nabla^{\alpha}\nabla_{\alpha'}
        \right]
        w(\mf x,\mf x')
    \right\}
    }
    \nonumber\\
    \tbb{
    =-\frac{1}{4\pi^2}\nabla_{\mu}[V_1].
    }
\end{align}
\tbb{Hence, the term $[V_1]g_{\mu\nu}/(4\pi^2)$ in
Eq.~\eqref{eq:renormalizedTmunu} is included to cancel this divergence
and restore local conservation.}

The remaining tensor
$\Theta_{\mu\nu}^{(L_{\mathrm{H}})}$ is a state-independent, local,
conserved tensor that encodes the finite renormalization freedom.
\tbb{In particular,}
\begin{equation}
    \nabla^{\nu}\Theta^{(L_{\mathrm{H}})}_{\mu\nu}=0.
\end{equation}
\tbb{Its addition preserves the conservation already established
by the first two terms in Eq.~\eqref{eq:renormalizedTmunu}:}
\begin{equation}
    \nabla^{\nu}
    \langle \mathopen{:}\hat{T}_{\mu\nu}\mathclose{:}\rangle
    =
    0.
\end{equation}
\tbb{The resulting expectation value can therefore serve as the
source in the semiclassical Einstein equations,}
\begin{equation}
    G_{\mu\nu}
    =
    8\pi
    \langle \mathopen{:}\hat{T}_{\mu\nu}\mathclose{:}\rangle.
    \label{eq:SCG}
\end{equation}

Different renormalization prescriptions satisfying the standard
requirements of locality, covariance, and conservation can differ by such a
state-independent conserved local curvature tensor
\cite{Wald1977BackReaction,Wald1978TraceAnomaly,bibleHadamard}.
\tbb{More explicitly, $\Theta_{\mu\nu}^{(L_{\mathrm{H}})}$ can be
written as a linear combination of the local conserved curvature
tensors allowed by these requirements, which must be fixed by additional renormalization conditions,
independently of the quantum state. 
In the semiclassical equations, this freedom can equivalently be
expressed as finite changes in the couplings multiplying local
curvature terms in the gravitational action
\cite{bibleHadamard,Hollands_Wald2001}.}

We now ask to what extent the renormalized source in
Eq.~\eqref{eq:SCG} can be expressed in terms of the
one-parameter family of UV-regulated Wightman functions.
\tbb{
The following derivation does not assume that
Eq.~\eqref{eq:SCG} holds. We work on a prescribed
background with a Hadamard state and reconstruct the renormalized
stress-energy expectation value independently of the Einstein
tensor. The semiclassical equation is an additional dynamical
condition relating these two quantities.}
Once the metric and its inverse have been reconstructed as in
Sec.~\ref{sec:main}, define the regulated point-splitting functional
\begin{equation}
    \mathcal{T}^{(\ell_{0})}_{\mu\nu}[W_{\ell_{0}}]
    \equiv
    [\partial_\mu\partial_{\nu'}W_{\ell_{0}}]
    -
    \frac{1}{2}g_{\mu\nu}g^{\alpha\beta}
    [\partial_\alpha\partial_{\beta'}W_{\ell_{0}}].
    \label{eq:regulated_stress_tensor_functional}
\end{equation}
For finite $\ell_{0}$, this quantity is well defined in a sufficiently small
neighbourhood of the diagonal. It should not, however, be interpreted as
the expectation value of a stress-energy operator. Since
$W_{\ell_{0}}$ need not be a physical two-point function of a
Klein--Gordon field, $\mathcal{T}^{(\ell_{0})}_{\mu\nu}[W_{\ell_{0}}]$ is only
an auxiliary regulated functional.

\tbb{Substituting the coincidence limit expansion
in Eq.~\eqref{eq:main_appendix}, with $m=\xi=0$, into
Eq.~\eqref{eq:regulated_stress_tensor_functional} gives}
\tbb{\begin{align}
    \mathcal{T}^{(\ell_0)}_{\mu\nu}[W_{\ell_0}]
    ={}&
    -\frac{g_{\mu\nu}}{2\pi^2\ell_0^4}
    -\frac{R_{\mu\nu}}{24\pi^2\ell_0^2}
    \nonumber\\
    &
    +\frac{1}{8\pi^2}
    \log\!\left(\frac{\ell_0^2}{2L_{\mathrm H}^2}\right)
    E_{\mu\nu}
    \nonumber\\
    &
    +\frac{1}{8\pi^2}
    \left(
        [w_{;\mu\nu'}]
        -\frac{1}{2}g_{\mu\nu}g^{\alpha\beta}
        [w_{;\alpha\beta'}]
    \right),
    \label{eq:regulated_stress_tensor_expansion}
\end{align}}
\tbb{where, for convenience, we defined}
\begin{equation}
    E_{\mu\nu}
    \equiv
    [V_{;\mu\nu'}]
    -
    \frac{1}{2}g_{\mu\nu}g^{\alpha\beta}
    [V_{;\alpha\beta'}].
    \label{eq:E_tensor}
\end{equation}
\tbb{In obtaining this expression, we used
$g^{\alpha\beta}g_{\alpha\beta}=4$ and
$g^{\alpha\beta}G_{\alpha\beta}=-R$.
Solving for the last line and substituting it into
Eq.~\eqref{eq:renormalizedTmunu}, including the conservation
term and finite renormalization freedom, yields}
\begin{align}
    \langle \mathopen{:}\hat{T}_{\mu\nu}\mathclose{:}\rangle
    ={}&
    \mathcal{T}^{(\ell_{0})}_{\mu\nu}[W_{\ell_{0}}]
    +
    \frac{g_{\mu\nu}}{2\pi^2\ell_{0}^4}
    +
    \frac{R_{\mu\nu}}{24\pi^2\ell_{0}^2}
    \nonumber\\
    &-
    \frac{1}{8\pi^2}
    \log\!\left(\frac{\ell_{0}^2}{2L_{\mathrm{H}}^2}\right)
    E_{\mu\nu}
    +
    \frac{[V_1]}{4\pi^2}g_{\mu\nu}
    +
    \Theta^{(L_{\mathrm{H}})}_{\mu\nu},
    \label{eq:regulated_semiclassical_source}
\end{align}
Equivalently, the renormalized stress-energy tensor is recovered by
removing the UV regulator:
\begin{align}
    \langle \mathopen{:}\hat{T}_{\mu\nu}\mathclose{:}\rangle
    ={}&
    \lim_{\ell_{0}\to0}
    \Biggl[
        \mathcal{T}^{(\ell_{0})}_{\mu\nu}[W_{\ell_{0}}]
        +
        \frac{g_{\mu\nu}}{2\pi^2\ell_{0}^4}
        +
        \frac{R_{\mu\nu}}{24\pi^2\ell_{0}^2}
        \nonumber\\
        &\hspace{1em}
        -
        \frac{1}{8\pi^2}
        \log\!\left(\frac{\ell_{0}^2}{2L_{\mathrm{H}}^2}\right)
        E_{\mu\nu}
    \Biggr]
    +
    \frac{[V_1]}{4\pi^2}g_{\mu\nu}
    +
    \Theta^{(L_{\mathrm{H}})}_{\mu\nu}.
    \label{eq:renormalized_stress_tensor_from_regulated_W}
\end{align}
\tbb{
The existence of this limit follows directly from
Eq.~\eqref{eq:regulated_stress_tensor_expansion}.
For the affine regularization used here,}
\tbb{\begin{align}
    &\mathcal{T}^{(\ell_0)}_{\mu\nu}[W_{\ell_0}]
    +\frac{g_{\mu\nu}}{2\pi^2\ell_0^4}
    +\frac{R_{\mu\nu}}{24\pi^2\ell_0^2}
    \nonumber\\
    &
    \qquad
    -\frac{1}{8\pi^2}
    \log\!\left(\frac{\ell_0^2}{2L_{\mathrm H}^2}\right)
    E_{\mu\nu}
    \nonumber\\
    &
    =
    \frac{1}{8\pi^2}
    \left(
        [w_{;\mu\nu'}]
        -\frac{1}{2}g_{\mu\nu}g^{\alpha\beta}
        [w_{;\alpha\beta'}]
    \right).
\end{align}}
\tbb{
Thus, the regulator-dependent terms cancel exactly for every
$\ell_0>0$. The right-hand side is independent of $\ell_0$
and finite because $w$ is smooth for a Hadamard state.
The limit therefore exists and reproduces the smooth-remainder
contribution to Eq.~\eqref{eq:renormalizedTmunu}.}

The functional
$\mathcal{T}^{(\ell_{0})}_{\mu\nu}[W_{\ell_{0}}]$, together with the
counterterms involving $g_{\mu\nu}$ and $R_{\mu\nu}$, is reconstructible
from the same one-parameter family of UV-regulated Wightman functions. \tbb{The same is true for the tensor $E_{\mu \nu}$, as we now show. For $\ell_{0} > 0$, introduce the auxiliary variable $s= \ell_{0}^2$. Then, a direct manipulation of Eq.~\eqref{eq:effective_metric_expansion} under the reparametrization $\ell_{0}^2 \to s$ yields}
\tbb{\begin{equation}
    [V_{;\mu \nu'}] = 4\pi^2 s \frac{d^3}{ds^3}\left(s^2\bigl[\partial_{\mu}\partial_{\nu'}W_{\sqrt{s}} \bigr]\right).
\end{equation}}
\tbb{Hence, it follows from the definition of $E_{\mu \nu}$ in Eq.~\eqref{eq:E_tensor} that}
\tbb{\begin{equation}
\begin{aligned}
    E_{\mu\nu}
    ={}& 4\pi^2 s \frac{d^3}{d s^3}
    \Bigl\{
        s^2
        \bigl[\partial_{\mu}\partial_{\nu'}W_{\sqrt{s}}\bigr]
    \\
    &\quad
        - \frac{s^2}{2}g_{\mu\nu}g^{\alpha\beta}
        \bigl[\partial_\alpha\partial_{\beta'}W_{\sqrt{s}}\bigr]
    \Bigr\}.
    \label{eq:E_from_regulated_W}
\end{aligned}
\end{equation}}

\tbb{The quantity $[V_1]$, on the other hand, involves derivatives of curvature as well as terms quadratic in curvature:}
\begin{equation}
    [V_1]
    =
    \frac{\Box R}{120}
    +
    \frac{R^2}{288}
    -
    \frac{R_{\alpha\beta}R^{\alpha\beta}}{720}
    +
    \frac{R_{\alpha\beta\mu\nu}R^{\alpha\beta\mu\nu}}{720},
    \label{eq:V1_minimal_massless}
\end{equation}
where
\begin{equation}
    \Box R
    =
    \frac{1}{\sqrt{-g}}
    \partial_\mu
    \left(
        \sqrt{-g}\,g^{\mu\nu}\partial_\nu R
    \right).
\end{equation}
\tbb{In principle, $[V_1]$ can be expressed entirely in terms of $W_{\ell_{0}}$, although we do not give the explicit expression here because of its complexity. Indeed, the last three terms can be reconstructed directly using the reconstruction formulas for the Ricci scalar and Ricci tensor, Eqs.~\eqref{eq:R_recover} and \eqref{eq:ricci_recover}, respectively, together with the formula for the Riemann tensor in Eq.~\eqref{eq:Riemann_from_W_l0}. The contractions can be performed using the inverse metric given in Eq.~\eqref{eq:inverse_metric_reconstruction}. The first term, $\Box R$, can likewise be expressed entirely in terms of $W_{\ell_{0}}$ via Eq.~\eqref{eq:volume_density_reconstruction}. To the best of our knowledge, no reconstruction formula for $[V_1]$ as compact as that for $E_{\mu\nu}$ is available. Nevertheless, the conceptual conclusion remains: $[V_1]$ can be reconstructed entirely from the UV-regulated Wightman function.}

\tbb{We next examine the dependence of the renormalization
prescription on the arbitrary Hadamard scale
$L_{\mathrm{H}}$. For a fixed geometry and quantum state,
changing $L_{\mathrm{H}}$ to $L_{\mathrm{H}}'$ modifies the
logarithmic term in the Hadamard representation. To leave
$W$ unchanged, the smooth remainder must transform as}
\begin{equation}
    \tbb{
    w \longrightarrow
    w+2\log\!\left(\frac{L_{\mathrm{H}}'}{L_{\mathrm{H}}}\right)V.
    }
\end{equation}
\tbb{Since $V$ is state-independent, this transformation
shifts the point-splitting contribution to the stress-energy
tensor by the local geometric term
$\log(L_{\mathrm{H}}'/L_{\mathrm{H}})E_{\mu\nu}/(4\pi^2)$.
Maintaining the same physical renormalization conditions
therefore requires the compensating transformation}
\begin{equation}
    \Theta_{\mu\nu}^{(L_{\mathrm{H}}')}
    =
    \Theta_{\mu\nu}^{(L_{\mathrm{H}})}
    -
    \frac{1}{4\pi^2}
    \log\!\left(\frac{L_{\mathrm{H}}'}{L_{\mathrm{H}}}\right)
    E_{\mu\nu}.
    \label{eq:consistency}
\end{equation}
\tbb{Thus, once the finite renormalization parameters
entering $\Theta_{\mu\nu}^{(L_{\mathrm{H}})}$ are specified
at a reference scale as inputs to the semiclassical model,
Eq.~\eqref{eq:consistency} determines their transformation
under a change of Hadamard scale. This ensures that the
complete renormalized source remains unchanged. Moreover,
Eq.~\eqref{eq:E_from_regulated_W} expresses the compensating
shift directly in terms of the regulated Wightman family.}

\tbb{In summary, we have shown that, once the Hadamard renormalization prescription is fixed, the expectation value of the stress-energy tensor can be expressed directly in terms of the UV-regulated Wightman function. Apart from the local ambiguity $\Theta^{(L_\mathrm{H})}_{\mu\nu}$, which must be supplied as an input to the model, every term in Eq.~\eqref{eq:regulated_semiclassical_source} admits a direct reconstruction from $W_{\ell_{0}}$. The previous section established that the same family of UV-regulated two-point functions also contains sufficient information to reconstruct the Einstein tensor. Thus, both sides of the semiclassical Einstein equations can be reconstructed independently from the same family of UV-regulated two-point functions. Imposing the semiclassical Einstein equations, Eq.~\eqref{eq:SCG}, then yields a set of highly nontrivial constraints on these two-point functions. Investigating these constraints and their implications for the interplay between quantum correlations and spacetime geometry is a natural direction for future work.}

\section{Conclusions and outlook}
\label{sec:conclusions}

The universal ultraviolet structure of quantum fields in curved spacetime encodes the local spacetime metric in field correlations \cite{Saravani_2016,Kempf_2021,Perche_Edu_geo_2022}. In this work, we investigated whether the same structure can be used to formulate the semiclassical Einstein equation in terms of a regulated family of field correlations. Such a formulation would sharpen the relation between quantum field theory and spacetime geometry and provide a concrete starting point for exploring emergent-gravity scenarios.

To this end, we introduced a local affine regularization of the ultraviolet structure of the Wightman function of a scalar quantum field, parametrized by an auxiliary regulator $\ell_{0}$. For a massless, minimally coupled scalar field, we derive, in the limit $\ell_{0}\to 0$, an expression for the Einstein tensor entirely in terms of this UV-regulated two-point structure. The same family of deformed correlations also furnishes direct reconstructions of the Ricci tensor, the Levi--Civita connection, and the Riemann tensor.

We also considered the matter side of semiclassical gravity. The regulated point-splitting functional constructed from $W_{\ell_{0}}$ reproduces the renormalized stress-energy tensor after the subtraction of universal regulator-dependent terms, up to the standard finite local renormalization freedom. Hence, once a prescription for this local ambiguity is specified, the geometric and matter contributions to the semiclassical Einstein equation can be organized in terms of the same UV-regulated correlation structure.

Taken together, these results sharpen a concrete question: can spacetime
geometry, rather than being introduced as independent background data,
arise from more fundamental quantum degrees of freedom and their
correlations?
\tbb{This question is also motivated by the work of Tjoa and
Gray~\cite{TjoaGray2022}, who reconstruct bulk geometry from
boundary field correlations, and, more broadly, by results
connecting entanglement equilibrium, relative entropy, and
spacetime thermodynamics to gravitational dynamics
\cite{Jacobson2016,Dorau2026,Serrano2025}.}

The present analysis does not provide such a mechanism.
The finite-$\ell_{0}$ object $W_{\ell_{0}}$ is an auxiliary
biscalar, and no consistent deformed quantum field theory is
presently known to realize it as a physical two-point function.

\tbb{A physical realization of $W_{\ell_{0}}$ at nonzero $\ell_{0}$
is not required for our reconstruction formulas, in which the
auxiliary regulator is ultimately removed. Nevertheless, an effective
theory that reproduces this biscalar, at least locally, as a physical
two-point function could provide a starting point for exploring an
operator-level interpretation of the reconstructed geometry.}
If such a realization were available, the relation in
Eq.~\eqref{eq:SCG} might admit a heuristic operator-level
reading: formally omitting the expectation value could suggest a
composite observable associated with the Einstein tensor.
Exploring the feasibility of such a construction, as well as
possible operational implementations of the UV regularization
described here, is left for future work.

More broadly, our results provide a correlation-based organization of the semiclassical Einstein equations and advance the programme of formulating spacetime geometry in terms of quantum field correlations, while opening new avenues for exploring a potential quantum emergence of spacetime geometry and gravitational dynamics.

\acknowledgements

AK acknowledges support through the Dieter Schwarz Foundation, the Natural Sciences and Engineering Research Council of Canada (NSERC), the National Research Council of Canada (NRC) and the Australian Research Council (ARC). MHZ thanks Prof. Achim Kempf for funding through his Dieter Schwarz grant. MHZ thanks Koji Yamaguchi for interesting discussions. Research at Perimeter Institute is supported in part by the Government of Canada through the Department of Innovation, Science and Industry Canada and by the Province of Ontario through the Ministry of Colleges and Universities.

\newpage

\onecolumngrid

\appendix

\section{A no-go result for direct Einstein-tensor reconstruction from the Wightman function}
\label{sec:proof_no_go}

In this appendix, we prove the claim made at the end of
Sec.~\ref{sec:metric_recovery}: within the class of direct constructions
that mirrors the metric-recovery prescription based on the Wightman
function, no universal pointwise transformation can yield the Einstein
tensor through a single mixed second coincidence derivative.

\tbb{
Our purpose is to determine whether replacing the scalar function
applied to the Wightman function, while retaining the derivative
structure of the metric recovery prescription, suffices to recover
the Einstein tensor. We restrict attention to this ansatz because
the higher-derivative route through the Riemann tensor (see Appendix~\ref{sec:riemann}) requires
additional contractions with the inverse metric to obtain the
Einstein tensor. The question is therefore whether a universal
pointwise transformation of the Wightman function can yield the
Einstein tensor directly through a single mixed second coincidence
derivative.}

Throughout this proof, $f(r)=\mathcal{O}(g(r))$ means that there exist constants $C>0$ and
$r_0>0$ such that $|f(r)|\leq C|g(r)|$ for $0<r<r_0$, whereas
$f(r)=o(g(r))$ means that $f(r)/g(r)\to0$ in the relevant limit.

In four spacetime dimensions, consider a Klein--Gordon field with parameters $m$ and $\xi$, and suppose, for the sake of contradiction, that there exists a universal function $ \Phi_{m,\xi}$ of class $C^{2}$ such that
\begin{equation}
    G_{\mu\nu}(\mf x)
    =
    \lim_{\mf x'\to\mf x}
    \partial_{\mu}\partial_{\nu'}
    \Phi_{m,\xi}\bigl(W(\mf x,\mf x')\bigr)
    \label{eq:no_go_appendix_Gmunu},
\end{equation}
for every Wightman function that obeys the Hadamard condition, as in Eq.~\eqref{eq:hadamard_first_time}. 

It is sufficient to consider the coincidence limit through spacelike
separations. Along such separations, the $\ii\epsilon$ prescription is
irrelevant, and the Hadamard singularity implies that the Wightman function
is real and tends to $+\infty$ as $\mf x'\to\mf x$. Consider first Minkowski
spacetime and assume that the field is prepared in the vacuum state. Let
\begin{equation}
    s\equiv \sigma_{0}(\mf x,\mf x')>0,
\end{equation}
where $s>0$ selects spacelike separated points. Moreover, the symmetries of Minkowski spacetime imply that the Wightman function depends only on $s$, so we write $W(s)$ in this case instead of $W(\mf x, \mf x')$. Then, by using that Hadamard form, it follows that
\begin{align}
    W(s)
    &=
    \frac{1}{8\pi^2s}
    +
    \mathcal{O}\!\left(
        \left|
        \log\!\left(\frac{s}{L_{\mathrm{H}}^{2}}\right)
        \right|
    \right),
    \label{eq:flat_wightman_asymptotics}
    \\
    W'(s)
    &=
    -\frac{1}{8\pi^2s^{2}}
    +
    \mathcal{O}\!\left(\frac{1}{s}\right),
    \label{eq:flat_wightman_first_derivative}
    \\
    W''(s)
    &=
    \frac{1}{4 \pi^2s^{3}}
    +
    \mathcal{O}\!\left(\frac{1}{s^{2}}\right).
    \label{eq:flat_wightman_second_derivative}
\end{align}
Now, define the auxiliary function
\begin{equation}
    h(s)
    \equiv
    \Phi_{m,\xi}\bigl(W(s)\bigr).
    \label{eq:definition_h_no_go}
\end{equation}
Writing $y^\mu=x^\mu-x'^\mu$, where $x^{\mu}$ are the coordinates associated with the point $\mf x$, the flat spacetime world function satisfies
\begin{equation}
    s
    =
    \frac{1}{2}\eta_{\alpha\beta}y^\alpha y^\beta,
    \qquad
    \partial_\mu s=y_\mu,
    \qquad
    \partial_{\nu'}s=-y_\nu,
    \qquad
    \partial_\mu\partial_{\nu'}s=-\eta_{\mu\nu}.
\end{equation}
Hence,
\begin{align}
    \partial_\mu\partial_{\nu'}
    \Phi_{m,\xi}\bigl(W(\mf x,\mf x')\bigr)
    &=
    \partial_\mu\partial_{\nu'}h(s)
    \nonumber\\
    &=
    -h'(s)\eta_{\mu\nu}
    -
    h''(s)y_\mu y_\nu.
    \label{eq:flat_mixed_derivative_no_go}
\end{align}
Since the Einstein tensor vanishes in Minkowski spacetime, the right-hand
side of Eq.~\eqref{eq:flat_mixed_derivative_no_go} must tend to zero as
$s\to0^{+}$. Choose coordinates such that
\begin{equation}
    y^\mu=(0,r,0,0),
    \qquad
    s=\frac{r^{2}}{2}.
\end{equation}
The $00$ component of Eq.~\eqref{eq:flat_mixed_derivative_no_go} then gives
\begin{equation}
    h'(s)=o(1)
    \qquad
    (s\to0^{+}),
    \label{eq:h_prime_vanishes}
\end{equation}
while the $11$ component gives
\begin{equation}
    -h'(s)-2s h''(s)=o(1).
\end{equation}
Together with Eq.~\eqref{eq:h_prime_vanishes}, this implies
\begin{equation}
    s\,h''(s)=o(1)
    \qquad
    (s\to0^{+}).
    \label{eq:s_h_double_prime_vanishes}
\end{equation}
We now translate these conditions into conditions on $\Phi_{m,\xi}$. First, we write
\begin{equation}
    h'(s)
    =
    \Phi_{m,\xi}'\bigl(W(s)\bigr)W'(s).
\end{equation}
Then, using Eqs.~\eqref{eq:flat_wightman_first_derivative} and
\eqref{eq:h_prime_vanishes}, one obtains
\begin{equation}
    \Phi_{m,\xi}'\bigl(W(s)\bigr)
    =
    o(s^{2}).
    \label{eq:phi_prime_small_s}
\end{equation}
Differentiating once more yields
\begin{equation}
    h''(s)
    =
    \Phi_{m,\xi}''\bigl(W(s)\bigr)
    \bigl(W'(s)\bigr)^2
    +
    \Phi_{m,\xi}'\bigl(W(s)\bigr)
    W''(s).
    \label{eq:second_chain_rule_no_go}
\end{equation}
The second term on the right-hand side is $o(s^{-1})$ by
Eqs.~\eqref{eq:flat_wightman_second_derivative} and
\eqref{eq:phi_prime_small_s}. Equation~\eqref{eq:s_h_double_prime_vanishes}
and $\bigl(W'(s)\bigr)^2\sim {1}/(64\pi^4s^{4})$ therefore imply
\begin{equation}
    \Phi_{m,\xi}''\bigl(W(s)\bigr)
    =
    o(s^{3}).
    \label{eq:phi_double_prime_small_s}
\end{equation}
For sufficiently small $s$, $W'(s)<0$, so $W$ is monotonic and its
range covers all sufficiently large positive values. Since
$W(s)\sim 1/(8\pi^2s)$, Eqs.~\eqref{eq:phi_prime_small_s} and
\eqref{eq:phi_double_prime_small_s} are equivalent to
\begin{equation}
    \Phi_{m,\xi}'(z)
    =
    o(z^{-2}),
    \qquad
    \Phi_{m,\xi}''(z)
    =
    o(z^{-3}),
    \qquad
    z\to+\infty.
    \label{eq:phi_large_argument_no_go}
\end{equation}
Under our assumption in Eq.~\eqref{eq:no_go_appendix_Gmunu}, the function $\Phi_{m,\xi}$ is universal and, therefore, the conditions in Eq.~\eqref{eq:phi_large_argument_no_go} must hold for any curved spacetime and any Hadamard state. In this general scenario, fix $\mf x$ and choose a unit spacelike tangent vector $u^\mu$ at
$\mf x$. For $ r>0$, define the point $\mf x'_{r}$ through the exponential map,
\begin{equation}
    \mf x'_r
    \equiv
    \exp_{\mf x}(r u).
\end{equation}
In this case,
\begin{equation}
    \sigma(\mf x,\mf x'_r)
    =
    \frac{r^{2}}{2},
    \label{eq:sigma_xr}
\end{equation}
and
\begin{equation}
    \Delta^{1/2}(\mf x,\mf x'_r)=1+\mathcal{O}(r^2).
    \label{eq:Deltar_leading}
\end{equation}
Then, using the Hadamard form, we can write, in any smooth coordinate
system,
\begin{equation}
    \partial_\mu W(\mf x,\mf x'_r)
    =
    \mathcal{O}(r^{-3}),
    \qquad
    \partial_{\nu'}W(\mf x,\mf x'_r)
    =
    \mathcal{O}(r^{-3}),
    \qquad
    \partial_\mu\partial_{\nu'}W(\mf x,\mf x'_r)
    =
    \mathcal{O}(r^{-4}).
    \label{eq:curved_wightman_derivative_estimates}
\end{equation}
For $r>0$ sufficiently small, define the real number
\begin{equation}
    W_r
    \equiv
    W(\mf x,\mf x'_r).
    \label{eq:definition_W_r_no_go}
\end{equation}
As explained above, the points $\mf x$ and $\mf x'_r$ are spacelike
separated. Hence $W_r$ is real and positive for sufficiently small $r$.
Using the Hadamard form together with
Eqs.~\eqref{eq:sigma_xr} and \eqref{eq:Deltar_leading}, we obtain
\begin{equation}
    W_r
    =
    \frac{1}{4\pi^2 r^2}
    +
    \mathcal{O}\!\left(
        \left|
        \log\!\left(\frac{r^2}{2L_{\mathrm{H}}^2}\right)
        \right|
    \right).
    \label{eq:curved_wightman_value_estimate}
\end{equation}
In particular, as $r \to 0^{+}$, we have
\begin{equation}
    W_r\to+\infty,
    \qquad
    W_r^{-2}
    =
    16\pi^4 r^4\bigl(1+o(1)\bigr),
    \qquad
    W_r^{-3}
    =
    64\pi^6 r^6\bigl(1+o(1)\bigr),
    \label{eq:inverse_W_r_estimates}
\end{equation}
Since $\Phi_{m,\xi}$ is the same universal function as in
the flat-spacetime argument, Eq.~\eqref{eq:phi_large_argument_no_go}
can be evaluated at $z=W_r$. It follows that
\begin{equation}
    \Phi_{m,\xi}'(W_r)
    =
    o(W_r^{-2})
    =
    o(r^4),
    \qquad
    \Phi_{m,\xi}^{''}(W_r)
    =
    o(W_r^{-3})
    =
    o(r^6).
    \label{eq:phi_curved_estimates_precise}
\end{equation}

For each fixed $r>0$, the Wightman function is smooth in a neighbourhood of
the separated pair $(\mf x,\mf x'_r)$. Since $\Phi_{m,\xi}$ is of class
$C^2$, the ordinary chain rule applies before the coincidence limit is
taken. Writing $\left. A\right|_r$ for the value of a biscalar or bitensor
$A(\mf x,\mf x')$ at $(\mf x,\mf x'_r)$, we have
\begin{equation}
    \left.
    \partial_\mu\partial_{\nu'}
    \Phi_{m,\xi}\bigl(W(\mf x,\mf x')\bigr)
    \right|_r=
    \Phi_{m,\xi}'(W_r)
    \left.
    \partial_\mu\partial_{\nu'}W(\mf x,\mf x')
    \right|_r
    +
     \Phi_{m,\xi}''(W_r)
    \left.
    \partial_\mu W(\mf x,\mf x')
    \right|_r
    \left.
    \partial_{\nu'}W(\mf x,\mf x')
    \right|_r.
    \label{eq:curved_chain_rule_no_go_precise}
\end{equation}
The estimates in Eq.~\eqref{eq:curved_wightman_derivative_estimates},
together with Eq.~\eqref{eq:phi_curved_estimates_precise}, imply
\begin{equation}
    \Phi_{m,\xi}'(W_r)
    \left.
    \partial_\mu\partial_{\nu'}W
    \right|_r=
    o(r^4)\mathcal{O}(r^{-4})
    =
    o(1),
\end{equation}
and
\begin{equation}
    \Phi_{m,\xi}''(W_r)
    \left.
    \partial_\mu W
    \right|_r
    \left.
    \partial_{\nu'}W
    \right|_r
    =
    o(r^6)\mathcal{O}(r^{-6})
    =
    o(1).
\end{equation}
Consequently,
\begin{equation}
    \lim_{r\to0^+}
    \left.
    \partial_\mu\partial_{\nu'}
    \Phi_{m,\xi}\bigl(W(\mf x,\mf x')\bigr)
    \right|_r
    =
    0.
    \label{eq:spacelike_limit_zero_no_go}
\end{equation}
Thus, the proposed expression has value zero along every fixed spacelike
approach to coincidence. If the coincidence limit in
Eq.~\eqref{eq:no_go_appendix_Gmunu} exists, it must agree with the
spacelike limit in Eq.~\eqref{eq:spacelike_limit_zero_no_go}. The assumed
construction would therefore imply $G_{\mu\nu}(\mf x)=0$ in every spacetime.
This contradicts the existence of spacetimes with nonvanishing Einstein
tensor that admit Hadamard states. Consequently, no universal function
$\Phi_{m,\xi}$ satisfying Eq.~\eqref{eq:no_go_appendix_Gmunu} exists.

\section{Effect of the UV regulator in momentum space}
\label{sec:fourier_stuff}

In this appendix, we show in detail how one can obtain Eq.~\eqref{eq:fourier_main_text} for the momentum-space representation of the regulated (massless) vacuum Wightman function in $3 + 1$ dimensions.

Let us start by defining the relative coordinates $\mf y \equiv \mf x - \mf x'$, so that $W_{\ell_{0}}$ can be written in terms of $\mf y$ (with $\mf y = (y^{\mu})$) as follows:
\begin{equation}
    W_{\ell_{0}}(\mf y) = \frac{1}{4\pi^2 (-(y^{0})^2 + |\boldsymbol{y}|^2 + L_{\mathrm{H}}^{2})}
    \label{eq:Wightmanl0mink_explicit}
\end{equation}
To evaluate the Fourier transform properly, we now restore the Wightman prescription. Thus, Eq.~\eqref{eq:Wightmanl0mink_explicit} should be understood as
\begin{equation}
    W_{\ell_{0}}(\mf y)
    =
    \frac{1}{4\pi^2}\,
    \frac{1}{-(y^{0}-\ii\epsilon)^2+|\boldsymbol{y}|^2+\ell_{0}^{2}},
    \qquad \epsilon\to 0^{+}.
    \label{eq:Wl0_lorentzian_ieps}
\end{equation}
It is convenient to split this expression as
\begin{equation}
    W_{\ell_{0}}(\mf y)
    =
    W_{0}(\mf y)+\Delta W_{\ell_{0}}(\mf y),
\end{equation}
where
\begin{equation}
    W_{0}(\mf y)
    =
    \frac{1}{4\pi^2}\frac{1}{-(y^{0}-\ii\epsilon)^2+|\boldsymbol{y}|^2},
\end{equation}
and
\begin{equation}
    \Delta W_{\ell_{0}}(\mf y)
    =
    \frac{1}{4\pi^2}
    \left[
        \frac{1}{-(y^{0}-\ii\epsilon)^2+|\boldsymbol{y}|^2+\ell_{0}^{2}}
        -
        \frac{1}{-(y^{0}-\ii\epsilon)^2+|\boldsymbol{y}|^2}
    \right].
    \label{eq:deltaW_correction}
\end{equation}
Let us now show that the Fourier transform of the first term is given by
\begin{equation}
    \widetilde{W}_{0}(\mf p)
    =
    2\pi\,\Theta(-p^{0})\,\delta(\mf p^{2}).
    \label{eq:FTW0}
\end{equation}
Let $\mf p = (\omega,\boldsymbol{p})$. Starting from
\begin{equation}
    W_{0}(\mf y)
    =
    \frac{1}{4\pi^2}\frac{1}{|\boldsymbol{y}|^2 - (t - \ii \epsilon)^2},
\end{equation}
we have
\begin{equation}
    \widetilde{W}_{0}(\mf p)
    =
    \frac{1}{4\pi^2}\int_{-\infty}^{\infty}\dd t \int_{\mathbb{R}^{3}} \dd^3\boldsymbol{y} \,
    \frac{e^{-\ii \omega t}e^{\ii \boldsymbol{p}\cdot \boldsymbol{y}}}{|\boldsymbol{y}|^2 - (t - \ii \epsilon)^2}.
\end{equation}
Let $r \equiv |\boldsymbol{y}|$. We first evaluate the integral over $t$:
\begin{equation}
    I_{\epsilon} = \int_{-\infty}^{\infty}\dd t \, f(t),
\end{equation}
where
\begin{equation}
    f(t) = \frac{e^{-\ii \omega t}}{r^2 - (t - \ii \epsilon)^2}.
\end{equation}
The poles are located at
\begin{equation}
    t_{1} = r + \ii \epsilon,
    \qquad
    t_{2} = -r + \ii \epsilon.
\end{equation}
Since $\epsilon > 0$, both poles lie in the upper half-plane. We now evaluate the integral by contour integration. One may close the contour with a semicircle of radius $R$ either in the lower or upper half-plane, and then take the limit $R \to \infty$. If $\omega > 0$, the exponential decays only in the lower half-plane, and so we close the contour there. Since there are no poles inside the contour, the integral vanishes. If $\omega < 0$, then the exponential decays in the upper half-plane, and we must close the contour there. In this case, both poles contribute, and the residue theorem yields
\begin{equation}
    I_{\epsilon}
    =
    \Theta(-\omega)\,2\pi \ii
    \left[
        \text{Res}(f,t=t_{1}) + \text{Res}(f,t=t_{2})
    \right]
    =
    \frac{2 \pi}{r}\Theta(-\omega)e^{\omega \epsilon}\sin(|\omega| r).
\end{equation}
Taking the limit $\epsilon \to 0^{+}$, we obtain
\begin{equation}
    \lim_{\epsilon \to 0^{+}}I_{\epsilon}
    =
    \frac{2 \pi}{r}\Theta(-\omega)\sin(|\omega| r).
\end{equation}
Substituting this back into the Fourier transform, we find
\begin{equation}
    \widetilde{W}_{0}(\mf p)
    =
    \frac{\Theta(-\omega)}{2\pi}
    \int_{\mathbb{R}^{3}} \dd^3\boldsymbol{y} \,
    e^{\ii \boldsymbol{p}\cdot \boldsymbol{y}}
    \frac{\sin(|\omega|r)}{r}.
\end{equation}
The remaining integral is spherically symmetric, so we may choose spherical coordinates with polar axis along $\boldsymbol{p}$. Then
\begin{align}
    \widetilde{W}_{0}(\mf p)
    &=
    \frac{\Theta(-\omega)}{2\pi}
    \int_{0}^{\infty}\dd r \, r^2 \frac{\sin(|\omega|r)}{r}
    \int_{0}^{2\pi}\dd\varphi
    \int_{0}^{\pi}\dd \theta \,\sin\theta \, e^{\ii |\boldsymbol{p}|r \cos\theta}
    \nonumber \\
    &=
    \frac{2 \Theta(-\omega)}{|\boldsymbol{p}|}
    \int_{0}^{\infty}\dd r \, \sin(|\omega|r)\sin(|\boldsymbol{p}|r).
\end{align}
To proceed, recall the distributional identity
\begin{equation}
    \delta(a) = \frac{1}{2\pi}\int_{-\infty}^{\infty}\dd x \, e^{\ii a x}.
\end{equation}
From it, one finds
\begin{equation}
    \int_{0}^{\infty}\dd x \, \sin(ax)\sin(bx)
    =
    \frac{\pi}{2}[\delta(a - b) - \delta(a + b)].
\end{equation}
Applying this result, and noting that $|\omega|$ and $|\boldsymbol{p}|$ are non-negative, we obtain
\begin{equation}
    \widetilde{W}_{0}(\mf p)
    =
    \pi \Theta(-\omega) \frac{\delta(|\omega| - |\boldsymbol{p}|)}{|\boldsymbol{p}|}.
\end{equation}
Finally, observe that
\begin{equation}
    \delta(\mf p^2)
    =
    \delta(-\omega^2 + |\boldsymbol{p}|^2)
    =
    \frac{\delta(|\omega| - |\boldsymbol{p}|)}{2|\boldsymbol{p}|}.
\end{equation}
Therefore,
\begin{equation}
    \widetilde{W}_{0}(\mf p)
    =
    2\pi \Theta(-p^{0})\delta(\mf p^2),
\end{equation}
\tbb{
which reproduces the standard massless vacuum result~\cite{PeskinSchroeder1995}}. Next, we compute the Fourier transform of the correction term. Writing
\begin{equation}
    \omega \equiv p^{0},
    \qquad
    k \equiv |\boldsymbol{p}|,
    \qquad
    r \equiv |\boldsymbol{y}|,
\end{equation}
we have
\begin{equation}
    \widetilde{\Delta W}_{\ell_{0}}(\mf p)
    =
    \frac{1}{4\pi^2}
    \int_{-\infty}^{\infty}\dd y^{0}\,e^{-\ii \omega y^{0}}
    \int_{\mathbb{R}^{3}}\dd^{3}\boldsymbol{y}\,
    e^{\ii \boldsymbol{p}\cdot\boldsymbol{y}}
    \left[
        \frac{1}{r^{2}+\ell_{0}^{2}-(y^{0}-\ii\epsilon)^2}
        -
        \frac{1}{r^{2}-(y^{0}-\ii\epsilon)^2}
    \right].
\end{equation}
\tbb{
The time integral is evaluated by the same contour argument
used above. For $a>0$, the integrand
$e^{-\ii\omega y^0}/[a^2-(y^0-\ii\epsilon)^2]$
has poles at $y^0=\pm a+\ii\epsilon$.
For $\omega>0$, we close the contour in the lower half-plane,
which contains no poles. For $\omega<0$, we close it in the
upper half-plane, where the sum of the residues is
$\ii e^{\omega\epsilon}\sin(\omega a)/a$. Thus,}
\begin{equation}
    \tbb{
    \int_{-\infty}^{\infty}\dd y^0\,
    \frac{e^{-\ii\omega y^0}}
         {a^2-(y^0-\ii\epsilon)^2}
    =
    -2\pi\Theta(-\omega)e^{\omega\epsilon}
    \frac{\sin(\omega a)}{a}.
    }
\end{equation}
\tbb{
Applying this formula with $a=\sqrt{r^2+\ell_0^2}$ and $a=r$,
with the latter expression understood by continuity at $r=0$,
and taking $\epsilon\to0^+$, we obtain}
\begin{align}
   & \int_{-\infty}^{\infty}\dd y^{0}\,e^{-\ii \omega y^{0}}
    \left[
        \frac{1}{r^{2}+\ell_{0}^{2}-(y^{0}-\ii\epsilon)^2}
        -
        \frac{1}{r^{2}-(y^{0}-\ii\epsilon)^2}
    \right]
     = \nonumber \\ & 
    -2\pi\,\Theta(-\omega)
    \left[
        \frac{\sin\!\bigl(\omega\sqrt{r^{2}+\ell_{0}^{2}}\bigr)}{\sqrt{r^{2}+\ell_{0}^{2}}}
        -
        \frac{\sin(\omega r)}{r}
    \right].
\end{align}
Thus,
\begin{equation}
    \widetilde{\Delta W}_{\ell_{0}}(\mf p)
    =
    -\frac{\Theta(-\omega)}{2\pi}
    \int_{\mathbb{R}^{3}}\dd^{3}\boldsymbol{y}\,
    e^{\ii \boldsymbol{p}\cdot\boldsymbol{y}}
    \left[
        \frac{\sin\!\bigl(\omega\sqrt{r^{2}+\ell_{0}^{2}}\bigr)}{\sqrt{r^{2}+\ell_{0}^{2}}}
        -
        \frac{\sin(\omega r)}{r}
    \right].
\end{equation}
The remaining integral is radial. Using\tbb{~\cite[Eq.~10.54.2]{DLMF} for $n=0$},
\begin{equation}
    \int_{\mathbb{R}^{3}}\dd^{3}\boldsymbol{y}\,
    e^{\ii \boldsymbol{p}\cdot\boldsymbol{y}}F(r)
    =
    4\pi \int_{0}^{\infty}\dd r\,r^{2}\,
    \frac{\sin(kr)}{kr}\,F(r),
\end{equation}
we obtain
\begin{equation}
    \widetilde{\Delta W}_{\ell_{0}}(\mf p)
    =
    -\frac{2\Theta(-\omega)}{k}
    \int_{0}^{\infty}\dd r\,
    \left[
        \frac{r\,\sin\!\bigl(\omega\sqrt{r^{2}+\ell_{0}^{2}}\bigr)}{\sqrt{r^{2}+\ell_{0}^{2}}}
        -
        \sin(\omega r)
    \right]
    \sin(kr).
    \label{eq:radial_integral_pre_bessel}
\end{equation}
\tbb{
To evaluate this integral, we use
$\partial_k\cos(kr)=-r\sin(kr)$.
We justify differentiating under the integral by first inserting
a damping factor $e^{-\eta r}$, with $\eta>0$, which makes both
the integral and its differentiated form absolutely convergent.
After differentiation, we remove the damping by taking
$\eta\to0^+$ in the distributional sense. Omitting this auxiliary
factor from the notation, we obtain}
\begin{align}
    & \int_{0}^{\infty}\dd r\,
    \Biggl[
        \frac{r\,\sin\!\bigl(\omega\sqrt{r^{2}+\ell_{0}^{2}}\bigr)}
             {\sqrt{r^{2}+\ell_{0}^{2}}}
        -
        \sin(\omega r)
    \Biggr]\sin(kr)
    \nonumber\\
    &\qquad =
    \frac{\partial}{\partial k}
    \int_0^\infty \dd r\,\cos(kr)
    \left[
        \frac{\sin(\omega r)}{r}
        -
        \frac{\sin\!\bigl(\omega\sqrt{\ell_0^{2}+r^2}\bigr)}
             {\sqrt{\ell_0^{2}+r^2}}
    \right].
\end{align}
As we are interested in the case $\omega < 0$, the following holds,
\begin{equation}
    \int_{0}^{\infty}\dd r\,
    \cos(kr)\frac{\sin(\omega r)}{r}
    =
    -\frac{\pi}{2}\Theta(-\omega-k).
\end{equation}
Moreover,\tbb{~\cite{GradshteynRyzhik2007}}
\begin{equation}
\int_0^\infty \dd r\,\cos(kr)\frac{\sin(\omega\sqrt{\ell^{2}_{0}+r^2})}{\sqrt{\ell^{2}_{0}+r^2}}
=
-\frac{\pi}{2}\,
J_0\!\left(\ell_{0}\sqrt{\omega^2-k^2}\right)\Theta(-\omega-k),
\end{equation}
where $J_{\alpha}(x)$ is the Bessel function of the first kind. Therefore,
\begin{align}
    & \int_{0}^{\infty}\dd r\,
    \Biggl[
        \frac{r\,\sin\!\bigl(\omega\sqrt{r^{2}+\ell_{0}^{2}}\bigr)}
             {\sqrt{r^{2}+\ell_{0}^{2}}}
        -
        \sin(\omega r)
    \Biggr]\sin(kr)
    \nonumber\\
    &\qquad =
    -\frac{\pi}{2}\frac{\partial}{\partial k}
    \left[
        \Bigl(
            1-
            J_0\!\bigl(\ell_0\sqrt{\omega^2-k^2}\bigr)
        \Bigr)
        \Theta(-\omega-k)
    \right].
\end{align}

Since $J_{0}(0)=1$, the derivative of the Heaviside factor produces no boundary contribution at $k=-\omega$. Hence,
\begin{equation}
    \int_{0}^{\infty}\dd r\,
    \Biggl[
        \frac{r\,\sin\!\bigl(\omega\sqrt{r^{2}+\ell_{0}^{2}}\bigr)}{\sqrt{r^{2}+\ell_{0}^{2}}}
        -
        \sin(\omega r)
    \Biggr]
    \sin(kr)
    =
    \frac{\pi \ell_{0} k}{2\sqrt{\omega^{2}-k^{2}}}
    J_{1}\!\bigl(\ell_{0}\sqrt{\omega^{2}-k^{2}}\bigr)
    \Theta(-\omega-k).
\end{equation}
Substituting this result into Eq.~\eqref{eq:radial_integral_pre_bessel}, we obtain
\begin{equation}
    \widetilde{\Delta W}_{\ell_{0}}(\mf p)
    =
    -\pi \ell_{0}\,
    \frac{
        J_{1}\!\bigl(\ell_{0}\sqrt{\omega^{2}-k^{2}}\bigr)
    }{
        \sqrt{\omega^{2}-k^{2}}
    }\,
    \Theta(-\omega)\Theta(-\omega-k).
\end{equation}
Since $k>0$, the factor $\Theta(-\omega-k)$ already implies $\omega<0$. Furthermore, this also implies that $\omega^2 - k^2 = -\mf p^2$ is positive, so the correction $\widetilde{\Delta W}_{\ell_{0}}(\mf p)$ is nonvanishing only for past-directed timelike four-momenta. Therefore, the Fourier transform of the UV-regulated Wightman function, Eq.~\eqref{eq:Wightmanl0mink_explicit}, can be written as
\begin{equation}
        \widetilde{W}_{\ell_{0}}(\mf p)  = 2\pi\,\Theta(-p^{0})\,\delta(\mf p^{2}) -\pi \ell_{0}\,
    \frac{
        J_{1}\!\bigl(\ell_{0}\sqrt{-\mf p^2}\bigr)
    }{
        \sqrt{-\mf p^2}
    }\,
    \Theta(-p^{0}-|\boldsymbol{p}|).
\end{equation}

\section{Explicit evaluation of the derivatives of the UV-regulated Wightman function}
\label{sec:calculations}

In this Appendix, we show how a general regularization of the UV structure of the Wightman function contains not only the metric, but also enough information to recover the Ricci and Einstein tensors. To that end, let us start by considering the general UV-regulated Wightman function
\begin{align}
W_{f}(\mf x,\mf x')
    & =
    \frac{1}{8\pi^2}
    \Biggl[
        \frac{\Delta^{1/2}(\mf x,\mf x')}{f(\sigma(\mf x, \mf x'))}+V(\mf x,\mf x')\log\!\left(\frac{f(\sigma(\mf x, \mf x'))}{L_{\mathrm{H}}^2}\right)
        +w(\mf x,\mf x')
    \Biggr],   \label{eq:hadamard_regulated_general}
\end{align}
where $f: \mathbb{R} \to \mathbb{R}$ is smooth, with $f(0) > 0$, and nonvanishing in a sufficiently small neighboorhood of $\sigma = 0$. In the main text, we focus on the choice $f(x) = x + \ell_{0}^2/2$.

Our goal is to evaluate the coincidence limit of $\nabla_{\mu}\nabla_{\nu'}W_{f}$. Let us start by differentiating with respect to $\mf x'$. We have
\begin{align}
\nabla_{\nu'}W_{f}(\mf x,\mf x')
&=
\frac{1}{8\pi^2}
\Biggl[
    \frac{(\Delta^{1/2}{})_{;\nu'}}{f}
    -
    \frac{\Delta^{1/2}f_{;\nu'}}{f^{2}}
    +
    V_{;\nu'}\log\!\left(\frac{f}{L_{\mathrm{H}}^2}\right)
    +
    \frac{Vf_{;\nu'}}{f}
    +
    w_{;\nu'}
\Biggr].
\end{align}
Differentiating once more with respect to $\mf x$, we find
\begin{align}
\nabla_{\mu}\nabla_{\nu'}W_{f}(\mf x,\mf x')
&=
\frac{1}{8\pi^2}
\Biggl[
    \frac{(\Delta^{1/2}{})_{;\mu\nu'}}{f}
    -
    \frac{
        (\Delta^{1/2}){}_{;\mu}f_{;\nu'}
        +
        (\Delta^{1/2}{})_{;\nu'}f_{;\mu}
        +
        \Delta^{1/2}f_{;\mu\nu'}
    }{
        f^{2}
    }
    \nonumber\\
&\qquad
    +
    2\frac{\Delta^{1/2}f_{;\mu}f_{;\nu'}}
    {f^{3}}
    +
    V_{;\mu\nu'}\log\!\left(\frac{f}{L_{\mathrm{H}}^2}\right)
    \nonumber\\
&\qquad
    +
    \frac{
        V_{;\mu}f_{;\nu'}
        +
        V_{;\nu'}f_{;\mu}
        +
        Vf_{;\mu\nu'}
    }{
        f} -
    \frac{Vf_{;\mu}f_{;\nu'}}
    {f^{2}}
    +
    w_{;\mu\nu'}
\Biggr].
\label{eq:second_derivative_Wl0}
\end{align}
Before evaluating the coincidence limit $\mf x' \to \mf x$, observe that $f_{;\mu} = f'(\sigma)\sigma_{;\mu}$, and $f_{;\mu \nu'} = f''(\sigma) \sigma_{;\mu}\sigma_{;\nu'} + f'(\sigma) \sigma_{;\mu \nu'}$. Hence, using the identities
\begin{equation}
       [\sigma_{;\mu}]
    =
    [\sigma_{;\nu'}]
    =
    0,
    \qquad
    [\sigma_{;\mu\nu'}]
    =
    -g_{\mu\nu},
\end{equation}
it follows that
\begin{equation}
    [f_{;\mu}] = 0, \qquad [f_{;\mu \nu'}] = -g_{\mu \nu}f'(0).
\end{equation}
Moreover, the coincidence limit of the van Vleck determinant and its derivatives reads \cite{poissonbible}
\begin{equation}
    [\Delta^{1/2}]
    =
    1,
    \qquad
    [\Delta^{1/2}{}_{;\mu}]
    =
    [\Delta^{1/2}{}_{;\nu'}]
    =
    0,
    \qquad
    [\Delta^{1/2}{}_{;\mu\nu'}]
    =
    -\frac{1}{6}R_{\mu\nu}.
\end{equation}
Therefore, the coincidence limit of $\nabla_{\mu}\nabla_{\nu'}W_{f}$ is given by
\begin{equation}
\bigl[\nabla_{\mu}\nabla_{\nu'}W_{f}\bigr] = \frac{1}{8\pi^2}\Biggl(\frac{f'(0)}{f(0)^2}g_{\mu \nu} - \frac{1}{6f(0)} R_{\mu \nu} - \frac{[V]f'(0)}{f(0)}g_{\mu \nu} + [V_{;\mu \nu'}]\log \left(\frac{f(0)}{L_{\mathrm{H}}^2}\right) + [w_{;\mu \nu'}]\Biggr).
\label{eq:jet_general}
\end{equation}
The coincidence limits $[V]$ and $[V_{;\mu \nu'}]$ can be explicitly written in terms of the local geometry and field parameters. In the case of $[V]$, we have \cite{bibleHadamard}
\begin{equation}
    [V] = \frac{m^2}{2} + \frac{R}{2}\left( \xi - \frac{1}{6}\right).
    \label{eq:V_coincidence}
\end{equation}
To write down an expression for $[V_{;\mu \nu'}]$, one can proceed as follows. Using the expansion in powers of Synge's world function,
\begin{equation}
    V(\mf x, \mf x')  = \sum_{n = 0}^{\infty}V_{n}(\mf x, \mf x')\sigma(\mf x, \mf x')^{n},
\end{equation}
it follows that
\begin{align}
    V_{;\mu \nu'}
    &=
    \sum_{n=0}^{\infty}
    \Bigl[
    (V_n)_{;\mu\nu'}\sigma^n
    +n(V_n)_{;\mu} \sigma^{n-1}\sigma_{;\nu'}
    +n(V_n)_{;\nu'} \sigma^{n-1}\sigma_{;\mu}
    +nV_n\sigma^{n-1}\sigma_{;\mu\nu'}
    \nonumber\\
    &\qquad\qquad
    +n(n-1)V_n\sigma^{n-2}\sigma_{;\mu}\sigma_{;\nu'}
    \Bigr].
\end{align}
Taking the coincidence limit, we find that all terms with $n\geq 2$ vanish. For $n=0$, only the term $(V_0)_{;\mu\nu'}$ survives, while for $n=1$ only the term $V_1\sigma_{;\mu\nu'}$ contributes. Therefore,
\begin{equation}
    [V_{;\mu\nu'}]
    =
    [V_{0\,;\mu\nu'}] + [V_1\sigma_{;\mu\nu'}]
    =
    [V_{0\,;\mu\nu'}] - [V_1]\,g_{\mu\nu}.
    \label{eq:coincidencehighermunuprime}
\end{equation}
The coincidence limit of the coefficient $V_{1}$ is given by \cite{bibleHadamard}:
\begin{equation}
    [V_1]
    =
    \frac{1}{8}m^4
    +\frac{1}{4}\left(\xi-\frac{1}{6}\right)m^2 R
    -\frac{1}{24}\left(\xi-\frac{1}{5}\right)\Box R +\frac{1}{8}\left(\xi-\frac{1}{6}\right)^2 R^2
    -\frac{1}{720}R_{\alpha\beta}R^{\alpha\beta}
    +\frac{1}{720}R_{\alpha\beta \mu \nu}R^{\alpha\beta \mu \nu}.
    \label{eq:V1_coincidence}
\end{equation}
As for $[V_{0 ;\mu \nu'}]$, a closed-form expression in terms of curvature tensors can be obtained using the covariant Taylor series of the biscalar $V_{0}(\mf x, \mf x')$. Indeed, we can write \cite{bibleHadamard}
\begin{equation}
    V_{0}(\mf x, \mf x') = [V_{0}(\mf x, \mf x')] + A_{\alpha}(\mf x)\sigma^{\alpha}(\mf x, \mf x') + \frac{1}{2}B_{\alpha\beta}(\mf x)\sigma^{\alpha}(\mf x, \mf x')\sigma^{\beta}(\mf x, \mf x') + \mathcal{O}(\sigma^{3/2})
\end{equation}
where
\begin{equation}
    A_{\alpha}
=
-\frac{1}{4}\left(\xi-\frac{1}{6}\right) R_{;\alpha}, \label{eq:Aalphageo}
\end{equation}
\begin{align}
B_{\alpha\beta}
&=
\frac{1}{12} m^{2} R_{\alpha\beta}
+\frac{1}{6}\left(\xi-\frac{3}{20}\right) R_{;\alpha\beta}
-\frac{1}{120}\Box R_{\alpha\beta}
+\frac{1}{12}\left(\xi-\frac{1}{6}\right) R\,R_{\alpha\beta}
\nonumber\\
&\quad
+\frac{1}{90} R^{\mu}{}_{\alpha} R_{\mu\beta}
-\frac{1}{180} R^{\mu\nu} R_{\mu\alpha\nu\beta}
-\frac{1}{180} R^{\mu\nu\rho}{}_{\alpha} R_{\mu\nu\rho\beta}.
\label{eq:Balphabetageo}
\end{align}
Hence,
\begin{align}
[V_{0;\mu\nu'}]
&=
-A_{\nu;\mu}
-
B_{\mu\nu}
\nonumber\\
&=
\frac{1}{4}\left(\xi-\frac{1}{6}\right)
R_{;\mu\nu}
-
B_{\mu\nu}
\nonumber\\
&=
\frac{1}{12}\left(\xi-\frac{1}{5}\right)
R_{;\mu\nu}
-\frac{1}{12}m^2R_{\mu\nu}
+\frac{1}{120}\Box R_{\mu\nu}
-\frac{1}{12}\left(\xi-\frac{1}{6}\right)R\,R_{\mu\nu}
\nonumber\\
&\quad
-\frac{1}{90}R^{\alpha}{}_{\mu}R_{\alpha\nu}
+\frac{1}{180}R^{\alpha\beta}R_{\alpha\mu\beta\nu}
+\frac{1}{180}R^{\alpha\beta\gamma}{}_{\mu}
R_{\alpha\beta\gamma\nu}.
\label{eq:coincidenciaV0munu}
\end{align}
Therefore, the coincidence limit $[V_{;\mu\nu'}]$ is a local tensor with higher curvature terms, explicitly determined by Eq.~\eqref{eq:coincidencehighermunuprime} together with Eqs.~\eqref{eq:V1_coincidence} and \eqref{eq:coincidenciaV0munu}. Having made the geometric content of all the terms in Eq.~\eqref{eq:jet_general} explicit, we can now identify how the different orders in the limit in which the regulator is removed encode different geometric information.

Equation~\eqref{eq:jet_general} shows that the leading and subleading terms in this limit are determined entirely by $f(0)$ and $f'(0)$. In particular, for a one-parameter family of regulators satisfying $f_{\ell_{0}}(0)\to 0$ and $f'_{\ell_{0}}(0)\to c\neq 0$, the leading divergence is proportional to the metric, while the subleading divergence contains the Ricci tensor. For the choice adopted in the main text,
\begin{equation}
    f(x)=x+\frac{\ell_{0}^2}{2},
\end{equation}
the general UV-regulated Wightman function $W_{f}$ reduces to $W_{\ell_{0}}$. In this case, we obtain
\begin{equation} 
[\nabla_{\mu}\nabla_{\nu'}W_{\ell_{0}}(\mf x,\mf x')]
=
\frac{g_{\mu\nu}}{2\pi^2\ell_{0}^4}
-
\frac{R_{\mu\nu}}{24\pi^2\ell_{0}^2}
- \frac{g_{\mu \nu}}{4\pi^2 \ell_{0}^2}\left([V] + \ell_{0}^2 \log\!\left(\frac{\ell_{0}^2}{2L_{\mathrm{H}}^2}\right) \frac{[V_1]}{2}\right) + \frac{[V_{0 \,;\mu \nu'}]}{8\pi^2}\log\!\left(\frac{\ell_{0}^2}{2L_{\mathrm{H}}^2}\right) +
\frac{[w_{;\mu\nu'}]}{8\pi^2}.
\label{eq:main_appendix}
\end{equation}

\section{Connection and curvature from the UV-regulated Wightman function}
\label{sec:riemann}

In principle, the Riemann curvature tensor can be written in terms of the Wightman function of a scalar field. As shown in Ref.~\cite{Perche_Edu_geo_2022}, one has
\begin{equation}
    R_{\alpha(\gamma\delta)\beta}
    =
    \frac{3}{4}
    \left(
        \frac{\Gamma\!\left(\frac{D}{2}-1\right)}
        {4\pi^{D/2}}
    \right)^{\frac{2}{D-2}}
    \left[
        \nabla_{\alpha}\partial_{\beta}
        \nabla_{\gamma'}\partial_{\delta'}
        \bigl(W(\mf x,\mf x')\bigr)^{\frac{2}{2-D}}
    \right].
\end{equation}
However, this expression does not yet provide a reconstruction solely in terms of the Wightman function, since the covariant derivatives on the right-hand side require knowledge of the connection and, therefore, of the spacetime metric.

This apparent circularity can be avoided because the connection can itself be reconstructed directly from the UV-regulated Wightman function considered here. To see this, recall that the coincidence limit of the third covariant derivative of Synge's world function vanishes:
\begin{equation}
    [\sigma_{;\lambda\mu'\nu'}]=0.
\end{equation}
Expressing the primed covariant derivatives in terms of partial derivatives gives
\begin{align}
    [\sigma_{;\lambda\mu'\nu'}]
    &=
    [\partial_{\lambda}\partial_{\mu'}\partial_{\nu'}\sigma]
    -
    \Gamma^{\rho}{}_{\mu\nu}
    [\partial_{\lambda}\partial_{\rho'}\sigma]
    \nonumber\\
    &=
    [\partial_{\lambda}\partial_{\mu'}\partial_{\nu'}\sigma]
    +
    \Gamma^{\rho}{}_{\mu\nu}g_{\lambda\rho}.
\end{align}
Then, it follows that
\begin{equation}
    [\partial_{\lambda}\partial_{\mu'}\partial_{\nu'}\sigma]
    =
    -\Gamma_{\lambda\mu\nu},
    \qquad
    \Gamma_{\lambda\mu\nu}
    \equiv
    g_{\lambda\rho}\Gamma^{\rho}{}_{\mu\nu}.
    \label{eq:sigma_connection}
\end{equation}

We now consider the third partial derivative of the general UV-regulated Wightman function introduced in Eq.~\eqref{eq:hadamard_regulated_general_main}. As the regulator is removed, the leading contribution is the term proportional to $f(0)^{-2}$. Thus,
\begin{align}
    \bigl[
        \partial_{\lambda}\partial_{\mu'}\partial_{\nu'}W_f
    \bigr]
    &=
    -\frac{1}{8\pi^2}
    \frac{f'(0)}{f(0)^2}
    [\partial_{\lambda}\partial_{\mu'}\partial_{\nu'}\sigma]
    +
    \mathcal{O}\!\left(\frac{1}{f(0)}\right)
    \nonumber\\
    &=
    \frac{1}{8\pi^2}
    \frac{f'(0)}{f(0)^2}
    \Gamma_{\lambda\mu\nu}
    +
    \mathcal{O}\!\left(\frac{1}{f(0)}\right).
    \label{eq:third_dev_general}
\end{align}
Therefore, for any one-parameter family of regulators satisfying
$f_{\ell_{0}}(0)\to0$ and $f'_{\ell_{0}}(0)\to c\neq0$, the connection with its first index lowered can be recovered as
\begin{equation}
    \Gamma_{\lambda\mu\nu}
    =
    8\pi^2
    \lim_{\ell_{0}\to0}
    \frac{f_{\ell_{0}}(0)^2}{f'_{\ell_{0}}(0)}
    \bigl[
        \partial_{\lambda}\partial_{\mu'}\partial_{\nu'}
        W_{f_{\ell_{0}}}(\mf x,\mf x')
    \bigr].
    \label{eq:connection_from_Wf}
\end{equation}
For the choice
$f_{\ell_{0}}(\sigma)=\sigma+\ell_{0}^2/2$, this becomes
\begin{equation}
    \Gamma_{\lambda\mu\nu}
    =
    \lim_{\ell_{0}\to0}
    2\pi^2\ell_{0}^4
    \bigl[
        \partial_{\lambda}\partial_{\mu'}\partial_{\nu'}
        W_{\ell_{0}}(\mf x,\mf x')
    \bigr].
    \label{eq:connection_lowered_from_W_l0}
\end{equation}
Finally, after using Eq.~\eqref{eq:inverse_metric_reconstruction} for the inverse metric reconstruction, we can raise the first index to obtain
\begin{equation}
    \Gamma^{\rho}{}_{\mu\nu}
    =
    g^{\rho\lambda}
    \lim_{\ell_{0}\to0}
    2\pi^2\ell_{0}^4
    \bigl[
        \partial_{\lambda}\partial_{\mu'}\partial_{\nu'}
        W_{\ell_{0}}(\mf x,\mf x')
    \bigr].
    \label{eq:connection_from_W_l0}
\end{equation}

With the connection in hand, we can proceed one step further and show that the Riemann tensor is also encoded in the UV-regulated Wightman function. We adopt the convention
\begin{equation}
    R^{\beta}{}_{\alpha\mu\nu}
    =
    \partial_{\mu}\Gamma^{\beta}{}_{\nu\alpha}
    -
    \partial_{\nu}\Gamma^{\beta}{}_{\mu\alpha}
    +
    \Gamma^{\beta}{}_{\mu\lambda}
    \Gamma^{\lambda}{}_{\nu\alpha}
    -
    \Gamma^{\beta}{}_{\nu\lambda}
    \Gamma^{\lambda}{}_{\mu\alpha}.
    \label{eq:Riemann_convention}
\end{equation}
Equivalently,
\begin{equation}
    R_{\beta\alpha\mu\nu}
    =
    \partial_{\mu}\Gamma_{\beta\nu\alpha}
    -
    \partial_{\nu}\Gamma_{\beta\mu\alpha}
    +
    \Gamma^{\lambda}{}_{\mu\alpha}\Gamma_{\lambda\nu\beta}
    -
    \Gamma^{\lambda}{}_{\nu\alpha}\Gamma_{\lambda\mu\beta}.
\end{equation}
To evaluate the derivatives of the coincidence limits, we use Synge's rule~\cite{poissonbible}. For any biscalar $B=B(\mf x,\mf x')$, it states that
\begin{equation}
    \partial_{\mu}[B]
    =
    [\partial_{\mu}B+\partial_{\mu'}B].
\end{equation}
Applying this rule to Eq.~\eqref{eq:connection_lowered_from_W_l0} gives
\begin{equation}
    \partial_{\mu}\Gamma_{\beta\nu\alpha}
    =
    2\pi^2
    \lim_{\ell_{0}\to0}
    \ell_{0}^4
    \bigl[
        \partial_{\mu}\partial_{\beta}
        \partial_{\nu'}\partial_{\alpha'}W_{\ell_{0}}
        +
        \partial_{\mu'}\partial_{\beta}
        \partial_{\nu'}\partial_{\alpha'}W_{\ell_{0}}
    \bigr].
\end{equation}
Because ordinary primed derivatives commute, the terms containing three primed derivatives cancel upon antisymmetrization under $\mu\leftrightarrow\nu$. We therefore obtain
\begin{align}
    R_{\beta\alpha\mu\nu}
    &=
    2\pi^2
    \lim_{\ell_{0}\to0}
    \ell_{0}^4
    \Bigl[
        \partial_{\mu}\partial_{\beta}
        \partial_{\nu'}\partial_{\alpha'}W_{\ell_{0}}
        -
        \partial_{\nu}\partial_{\beta}
        \partial_{\mu'}\partial_{\alpha'}W_{\ell_{0}}
    \Bigr]
    \nonumber\\
    &\qquad
    +
    \Gamma^{\lambda}{}_{\mu\alpha}
    \Gamma_{\lambda\nu\beta}
    -
    \Gamma^{\lambda}{}_{\nu\alpha}
    \Gamma_{\lambda\mu\beta}.
    \label{eq:Riemann_from_fourth_derivatives_intermediate}
\end{align}
Substituting the reconstruction formulas for the connection and inverse metric then yields an expression entirely in terms of $W_{\ell_{0}}$:
\begin{align}
    R_{\beta\alpha\mu\nu}
    &=
    2\pi^2
    \lim_{\ell_{0}\to0}
    \ell_{0}^4
    \Biggl\{
        \Bigl[
            \partial_{\mu}\partial_{\beta}
            \partial_{\nu'}\partial_{\alpha'}W_{\ell_{0}}
            -
            \partial_{\nu}\partial_{\beta}
            \partial_{\mu'}\partial_{\alpha'}W_{\ell_{0}}
        \Bigr]
        \nonumber\\
    &+
        \left(
            \bigl[\partial\partial'W_{\ell_{0}}\bigr]^{-1}
        \right)^{\lambda\kappa}
        \Bigl(
            \bigl[
                \partial_{\kappa}\partial_{\mu'}\partial_{\alpha'}
                W_{\ell_{0}}
            \bigr]
            \bigl[
                \partial_{\lambda}\partial_{\nu'}\partial_{\beta'}
                W_{\ell_{0}}
            \bigr]
        \nonumber\\
    &\qquad\qquad
            -
            \bigl[
                \partial_{\kappa}\partial_{\nu'}\partial_{\alpha'}
                W_{\ell_{0}}
            \bigr]
            \bigl[
                \partial_{\lambda}\partial_{\mu'}\partial_{\beta'}
                W_{\ell_{0}}
            \bigr]
        \Bigr)
    \Biggr\}.
    \label{eq:Riemann_from_W_l0}
\end{align}

\twocolumngrid

\bibliography{references}

@article{Perche_Edu_geo_2022,
   title={Geometry of spacetime from quantum measurements},
   volume={105},
   ISSN={2470-0029},
   url={http://dx.doi.org/10.1103/PhysRevD.105.066011},
   number={6},
   journal={Phys. Rev. D},
   publisher={American Physical Society (APS)},
   author={Perche, T. Rick and Martín-Martínez, Eduardo},
   year={2022},
   month=mar }

@article{Kempf_2021,
   title={Replacing the Notion of Spacetime Distance by the Notion of Correlation},
   volume={9},
   ISSN={2296-424X},
   url={http://dx.doi.org/10.3389/fphy.2021.655857},
   journal={Front. Phys.},
   publisher={Frontiers Media SA},
   author={Kempf, Achim},
   year={2021},
   month=may }

@article{AhmedMatheusTales2025,
	author={Shalabi, Ahmed and Hrabowec Zambianco, Matheus and Rick Perche, Tales},
	title={The effect of curvature on local observables in quantum field theory},
	journal={Classical and Quantum Gravity},
	url={http://iopscience.iop.org/article/10.1088/1361-6382/adc534},
	year={2025}
}

@article{poissonbible,
   title={The Motion of Point Particles in Curved Spacetime},
   volume={7},
   ISSN={1433-8351},
   url={http://dx.doi.org/10.12942/lrr-2004-6},
   number={1},
   journal={Living Rev. Relativ.},
   publisher={Springer Nature},
   author={Poisson, Eric},
   year={2004},
   month={May}
}

@book{PeskinSchroeder1995,
  author    = {Peskin, Michael E. and Schroeder, Daniel V.},
  title     = {An Introduction to Quantum Field Theory},
  publisher = {Addison-Wesley},
  address   = {Reading, Massachusetts},
  year      = {1995},
  isbn      = {9780201503975}
}

@book{GradshteynRyzhik2007,
  author    = {Gradshteyn, I. S. and Ryzhik, I. M.},
  title     = {Table of Integrals, Series, and Products},
  editor    = {Jeffrey, Alan and Zwillinger, Daniel},
  edition   = {7},
  publisher = {Academic Press},
  year      = {2007},
  isbn      = {9780123736376}
}

@misc{DLMF,
  title        = {{NIST Digital Library of Mathematical Functions}},
  organization = {National Institute of Standards and Technology},
  url          = {https://dlmf.nist.gov/10.54.E2},
  note         = {Eq.~10.54.2}
}

@article{anderson2003,
  title = {Linear response, validity of semiclassical gravity, and the stability of flat space},
  author = {Anderson, Paul R. and Molina-Par{\'{\i}}s, Carmen and Mottola, Emil},
  journal = {Phys. Rev. D},
  volume = {67},
  issue = {2},
  pages = {024026},
  numpages = {19},
  year = {2003},
  month = {Jan},
  publisher = {American Physical Society},
  url = {https://link.aps.org/doi/10.1103/PhysRevD.67.024026}
}

@article{Huetal2004,
  title = {Induced quantum metric fluctuations and the validity of semiclassical gravity},
  author = {Hu, B. L. and Roura, Albert and Verdaguer, Enric},
  journal = {Phys. Rev. D},
  volume = {70},
  issue = {4},
  pages = {044002},
  numpages = {24},
  year = {2004},
  month = {Aug},
  publisher = {American Physical Society},
  url = {https://link.aps.org/doi/10.1103/PhysRevD.70.044002}
}

@article{HuVerdaguer2008,
  author  = {Hu, Bei Lok and Verdaguer, Enric},
  title   = {Stochastic Gravity: Theory and Applications},
  journal = {Living Rev. Relativity},
  volume  = {11},
  pages   = {3},
  year    = {2008},
  eprint  = {0802.0658},
}

@article{Moretti2021GlobalHadamard,
  author  = {Moretti, Valter},
  title   = {On the global {Hadamard} parametrix in {QFT} and the
             signed squared geodesic distance defined in domains
             larger than convex normal neighbourhoods},
  journal = {Letters in Mathematical Physics},
  volume  = {111},
  pages   = {130},
  year    = {2021},
  doi     = {10.1007/s11005-021-01464-4}
}

@article{Unruh1976,
  title = {Notes on black-hole evaporation},
  author = {Unruh, W. G.},
  journal = {Phys. Rev. D},
  volume = {14},
  issue = {4},
  pages = {870--892},
  numpages = {0},
  year = {1976},
  month = {Aug},
  publisher = {American Physical Society},
  doi = {10.1103/PhysRevD.14.870},
  url = {https://link.aps.org/doi/10.1103/PhysRevD.14.870}
}

@article{fewsterNecessityHadamard,
	doi = {10.1088/0264-9381/30/23/235027},
	url = {https://doi.org/10.1088/0264-9381/30/23/235027},
	year = 2013,
	month = {nov},
	publisher = {{IOP} Publishing},
	volume = {30},
	number = {23},
	pages = {235027},
	author = {Christopher J Fewster and Rainer Verch},
	title = {The necessity of the Hadamard condition},
	journal = {Class. Quantum Grav. }
}

@book{ParkerToms,
    author    = {Leonard Parker and David J. Toms},
    title     = {Quantum Field Theory in Curved Spacetime: Quantized Fields and Gravity},
    publisher = {Cambridge University Press},
    address   = {Cambridge},
    year      = {2009},
    doi       = {10.1017/CBO9780511813924}
}

@article{Christensen1976,
    author  = {S. M. Christensen},
    title   = {Vacuum Expectation Value of the Stress Tensor in an Arbitrary Curved Background: The Covariant Point-Separation Method},
    journal = {Phys. Rev. D},
    volume  = {14},
    pages   = {2490--2501},
    year    = {1976},
    doi     = {10.1103/PhysRevD.14.2490}
}

@article{Fulling1973,
    author  = {Stephen A. Fulling},
    title   = {Nonuniqueness of Canonical Field Quantization in Riemannian Space-Time},
    journal = {Phys. Rev. D},
    volume  = {7},
    pages   = {2850--2862},
    year    = {1973},
    doi     = {10.1103/PhysRevD.7.2850}
}

@article{BardeenCarterHawking1973,
    author  = {James M. Bardeen and Brandon Carter and Stephen W. Hawking},
    title   = {The Four Laws of Black Hole Mechanics},
    journal = {Commun. Math. Phys.},
    volume  = {31},
    pages   = {161--170},
    year    = {1973},
    doi     = {10.1007/BF01645742}
}

@article{Parker1968,
    author  = {Leonard Parker},
    title   = {Particle Creation in Expanding Universes},
    journal = {Phys. Rev. Letters},
    volume  = {21},
    pages   = {562--564},
    year    = {1968},
}

@book{WaldQFTCS,
    author    = {Robert M. Wald},
    title     = {Quantum Field Theory in Curved Spacetime and Black Hole Thermodynamics},
    publisher = {University of Chicago Press},
    address   = {Chicago},
    year      = {1994}
}

@book{BirrellDavies,
    author    = {N. D. Birrell and P. C. W. Davies},
    title     = {Quantum Fields in Curved Space},
    publisher = {Cambridge University Press},
    address   = {Cambridge},
    year      = {1982},
}

@article{tHooftVeltman1972,
title = {Regularization and renormalization of gauge fields},
journal = {Nucl. Phys. B},
volume = {44},
number = {1},
pages = {189-213},
year = {1972},
issn = {0550-3213},
url = {https://www.sciencedirect.com/science/article/pii/0550321372902799},
author = {G. {'t Hooft} and M. Veltman},
}

@article{Bollini1972,
  author  = {Bollini, C. G. and Giambiagi, J. J.},
  title   = {Dimensional renorinalization: The number of dimensions as a regularizing parameter},
  journal = {Nuovo Cim. B},
  volume  = {12},
  number  = {1},
  pages   = {20--26},
  year    = {1972},
  url     = {https://doi.org/10.1007/BF02895558}
}

@article{Hollands_Wald2001,
   title={Local Wick Polynomials and Time Ordered Products of Quantum Fields in Curved Spacetime},
   volume={223},
   ISSN={1432-0916},
   url={http://dx.doi.org/10.1007/s002200100540},
   number={2},
   journal={Commun. Math. Phys.},
   publisher={Springer Science and Business Media LLC},
   author={Hollands, Stefan and Wald, Robert M.},
   year={2001},
   month=Oct, pages={289–326} }

@article{Padmanabhan_1997,
  author  = {Padmanabhan, T.},
  title   = {Duality and Zero-Point Length of Spacetime},
  journal = {Phys. Rev. Letters},
  volume  = {78},
  number  = {10},
  pages   = {1854--1857},
  year    = {1997},
}

@article{Radzikowski_1996,
  author  = {Radzikowski, Marek J.},
  title   = {Micro-Local Approach to the Hadamard Condition in Quantum Field Theory on Curved Space-Time},
  journal = {Commun. Math. Phys.},
  volume  = {179},
  pages   = {529--553},
  year    = {1996},
  doi     = {10.1007/BF02100096}
}

@article{GibbonsHawking_1977,
  title = {Cosmological event horizons, thermodynamics, and particle creation},
  author = {Gibbons, G. W. and Hawking, S. W.},
  journal = {Phys. Rev. D},
  volume = {15},
  issue = {10},
  pages = {2738--2751},
  numpages = {0},
  year = {1977},
  month = {May},
  publisher = {American Physical Society},
  doi = {10.1103/PhysRevD.15.2738},
  url = {https://link.aps.org/doi/10.1103/PhysRevD.15.2738}
}

@article{fullingHadamard2,
    author = "Fulling, S. A. and Narcowich, F. J. and Wald, Robert M.",
    title = "{Singularity Structure of the Two Point Function in Quantum Field Theory in Curved Space-time. {II}}",
    doi = "10.1016/0003-4916(81)90098-1",
    journal = "Annals Phys.",
    volume = "136",
    pages = "243--272",
    year = "1981"
}

@Article{fullingHadamard,
author={Fulling, Stephen A.
and Sweeny, Mark
and Wald, Robert M.},
title={Singularity structure of the two-point function in quantum field theory in curved spacetime},
journal={Commun. Math. Phys},
year={1978},
month={Oct},
day={01},
volume={63},
number={3},
pages={257-264},
issn={1432-0916},
doi={10.1007/BF01196934},
url={https://doi.org/10.1007/BF01196934}
}

@article{Dorau2026,
   title={From Quantum Relative Entropy to the Semiclassical Einstein Equations},
   volume={136},
   ISSN={1079-7114},
   url={http://dx.doi.org/10.1103/lmq8-nsty},
   number={9},
   journal={Phys. Rev. Lett.},
   publisher={American Physical Society (APS)},
   author={Dorau, Philipp and Much, Albert},
   year={2026},
   month=Mar }

@article{Serrano2025,
  title = {Gravity from equilibrium thermodynamics of stretched light cones},
  author = {Alonso-Serrano, Ana and Garay, Luis J. and Li{\ifmmode \check{s}\else \v{s}\fi{}}ka, Marek and L{\'o}pez Pineros, Celia},
  journal = {Phys. Rev. D},
  volume = {112},
  issue = {12},
  pages = {124026},
  numpages = {14},
  year = {2025},
  month = {Dec},
  publisher = {American Physical Society},
  url = {https://link.aps.org/doi/10.1103/y9rz-b88v}
}

@article{Jacobson2016,
  title = {Entanglement Equilibrium and the Einstein Equation},
  author = {Jacobson, Ted},
  journal = {Phys. Rev. Lett.},
  volume = {116},
  issue = {20},
  pages = {201101},
  numpages = {6},
  year = {2016},
  month = {May},
  publisher = {American Physical Society},
  url = {https://link.aps.org/doi/10.1103/PhysRevLett.116.201101}
}

@article{TjoaGray2022,
  title = {Modest holography and bulk reconstruction in asymptotically flat spacetimes},
  author = {Tjoa, Erickson and Gray, Finnian},
  journal = {Phys. Rev. D},
  volume = {106},
  issue = {2},
  pages = {025021},
  numpages = {24},
  year = {2022},
  month = {Jul},
  publisher = {American Physical Society},
  url = {https://link.aps.org/doi/10.1103/PhysRevD.106.025021}
}

@article{matsasUnruh,
  title = {The {Unruh} effect and its applications},
  author = {Crispino, Lu\'{\i}s C. B. and Higuchi, Atsushi and Matsas, George E. A.},
  journal = {Rev. Mod. Phys.},
  volume = {80},
  issue = {3},
  pages = {787--838},
  numpages = {0},
  year = {2008},
  month = {Jul},
  publisher = {American Physical Society},
  doi = {10.1103/RevModPhys.80.787},
  url = {https://link.aps.org/doi/10.1103/RevModPhys.80.787}
}

@article{HawkingRadiation,
    author = "Hawking, S. W.",
    title = "{Black hole explosions}",
    doi = "10.1038/248030a0",
    journal = "Nature",
    volume = "248",
    pages = "30--31",
    year = "1974"
}

@article{Bekenstein1973,
    author  = {Jacob D. Bekenstein},
    title   = {Black Holes and Entropy},
    journal = {Phys. Rev. D},
    volume  = {7},
    pages   = {2333--2346},
    year    = {1973},
}

@article{ParkerOG,
  title = {Quantized Fields and Particle Creation in Expanding Universes. I},
  author = {Parker, Leonard},
  journal = {Phys. Rev.},
  volume = {183},
  issue = {5},
  pages = {1057--1068},
  numpages = {0},
  year = {1969},
  month = {Jul},
  publisher = {American Physical Society},
  doi = {10.1103/PhysRev.183.1057},
  url = {https://link.aps.org/doi/10.1103/PhysRev.183.1057}
}

@Article{Hawking1975,
author={Hawking, S. W.},
title={Particle creation by black holes},
journal={Commun. Math. Phys.},
year={1975},
month={Aug},
day={01},
volume={43},
number={3},
pages={199-220},
issn={1432-0916},
doi={10.1007/BF02345020},
url={https://doi.org/10.1007/BF02345020}
}

@article{bibleHadamard,
   title={Hadamard renormalization of the stress-energy tensor for a quantized scalar field in a general spacetime of arbitrary dimension},
   volume={78},
   ISSN={1550-2368},
   url={http://dx.doi.org/10.1103/PhysRevD.78.044025},
   journal={Phys. Rev. D},
   publisher={American Physical Society (APS)},
   author={Décanini, Yves and Folacci, Antoine},
   year={2008},
   month=aug }

@article{Stargen_2015,
   title={Small scale structure of spacetime: The van Vleck determinant and equigeodesic surfaces},
   volume={92},
   ISSN={1550-2368},
   url={http://dx.doi.org/10.1103/PhysRevD.92.024046},
   number={2},
   journal={Phys. Rev. D},
   publisher={American Physical Society (APS)},
   author={Stargen, D. Jaffino and Kothawala, Dawood},
   year={2015},
   month=jul }

@article{Kothawala_2013,
   title={Minimal length and small scale structure of spacetime},
   volume={88},
   ISSN={1550-2368},
   url={http://dx.doi.org/10.1103/PhysRevD.88.104029},
   number={10},
   journal={Phys. Rev. D},
   publisher={American Physical Society (APS)},
   author={Kothawala, Dawood},
   year={2013},
   month=nov }

@article{Perche_Ahmed_2022,
   title={Spacetime curvature from ultrarapid measurements of quantum fields},
   volume={105},
   ISSN={2470-0029},
   url={http://dx.doi.org/10.1103/PhysRevD.105.125011},
   number={12},
   journal={Phys. Rev. D},
   publisher={American Physical Society (APS)},
   author={Perche, T. Rick and Shalabi, Ahmed},
   year={2022},
   month=jun }

@article{Wald1977BackReaction,
  author  = {Wald, Robert M.},
  title   = {The Back Reaction Effect in Particle Creation in Curved Space-Time},
  journal = {Commun. Math. Phys.},
  volume  = {54},
  pages   = {1--19},
  year    = {1977},
  doi     = {10.1007/BF01609833}
}

@article{Wald1978TraceAnomaly,
  author  = {Wald, Robert M.},
  title   = {Trace Anomaly of a Conformally Invariant Quantum Field in Curved Space-Time},
  journal = {Phys. Rev. D},
  volume  = {17},
  pages   = {1477--1484},
  year    = {1978},
  doi     = {10.1103/PhysRevD.17.1477}
}

@article{Saravani_2016,
   title={Spacetime curvature in terms of scalar field propagators},
   volume={93},
   ISSN={2470-0029},
   url={http://dx.doi.org/10.1103/PhysRevD.93.045026},
   number={4},
   journal={Phys. Rev. D},
   publisher={American Physical Society (APS)},
   author={Saravani, Mehdi and Aslanbeigi, Siavash and Kempf, Achim},
   year={2016},
   month=Feb }

@article{DeWitt81,
  title = {Approximate Effective Action for Quantum Gravity},
  author = {DeWitt, Bryce S.},
  journal = {Phys. Rev. Lett.},
  volume = {47},
  issue = {23},
  pages = {1647--1650},
  numpages = {0},
  year = {1981},
  month = {Dec},
  publisher = {American Physical Society},
  doi = {10.1103/PhysRevLett.47.1647},
  url = {https://link.aps.org/doi/10.1103/PhysRevLett.47.1647}
}

@inbook{Khavkine_2015,
   title={Algebraic QFT in Curved Spacetime and Quasifree Hadamard States: An Introduction},
   ISBN={9783319213538},
   ISSN={2352-3905},
   url={http://dx.doi.org/10.1007/978-3-319-21353-8_5},
   DOI={10.1007/978-3-319-21353-8_5},
   booktitle={Advances in Algebraic Quantum Field Theory},
   publisher={Springer International Publishing},
   author={Khavkine, Igor and Moretti, Valter},
   year={2015},
   pages={191–251} }

\end{document}